\documentclass[pra,aps,twocolumn,superscriptaddress]{revtex4-2}

\usepackage{amssymb}
\usepackage{amsmath}
\usepackage{amsfonts}
\usepackage{bbold}
\usepackage{xfrac}

\usepackage[T1]{fontenc}
\usepackage[utf8]{inputenc}
\usepackage{epsfig,pstricks,graphicx}
\usepackage{bm}

\usepackage{color}
\def\id{{\rm 1\kern-.22em l}}
\usepackage{enumerate}
\usepackage[capitalize]{cleveref}

\usepackage{float}
\usepackage{soul}

\usepackage{lipsum}
\usepackage{multirow}
\usepackage{systeme}

\begin{document}

\title{Quantum Correlations in Three-Beam Symmetric Gaussian States Accessed via Photon-Number-Resolving Detection and Quantum Universal Invariants}

\author{Jan Pe\v{r}ina Jr.}
\email{jan.perina.jr@upol.cz}
\affiliation{Joint Laboratory of Optics of Palack\'{y} University and Institute of Physics of CAS, Faculty of Science, Palack\'{y} University, 17. listopadu 12, 771 46 Olomouc, Czech Republic}
\author{Nazarii Sudak}
\affiliation{Institute of Theoretical Physics, University of
Wroclaw, Plac Maxa Borna 9, 50-204 Wrocław, Poland}
\author{Artur Barasi\'nski}
\email{artur.barasinski@uwr.edu.pl}
\affiliation{Institute of Theoretical Physics, University of Wroclaw, Plac Maxa Borna 9, 50-204 Wrocław, Poland}
\author{Anton\' in \v{C}ernoch}
\affiliation{Joint Laboratory of Optics, Institute of Physics of CAS, 17. listopadu 50a, 779 00 Olomouc, Czech Republic}

\begin{abstract}
Quantum correlations of symmetric 3-beam Gaussian states are analyzed using their quantum universal invariants. These invariants, 1-, 2-, and 3-beam purities, are expressed in terms of the beams' intensity moments up to sixth order. The 3-beam symmetric Gaussian states with varying amounts of the noise are experimentally generated
using entangled photon pairs from down-conversion, their invariants are determined, and their quantum correlations are quantified.
The coexistence of bi- and tripartite entanglement and genuine tripartite entanglement is observed in these states that resemble the noisy GHZ/W states.
\end{abstract}

\maketitle

\section{Introduction}

The discovery of phase-squeezed light~\cite{Stoler1970}, whose
observation~\cite{Slusher1985} belongs to the first experiments
certifying the existence of nonclassical states of light,
triggered fast development of the wide area of quantum optics
devoted to the properties of bosonic modes with their infinitely
large Hilbert spaces~\cite{Dodonov2002}. Today, this area, known
as continuous-variable (CV) quantum optics, represents a
particularly promising avenue to develop different types of
quantum technologies applicable, e.g., in quantum
information processing and quantum key distribution (QKD)~\cite{Laudenbach2018}. Compared to
other developing quantum technologies (e.g., based on discrete
variables), CV states exhibit resilience against decoherence and
provide large Hilbert spaces for information encoding. They also
allow for purification~\cite{Franzen2006} and entanglement
distillation~\cite{Grebien2022}. Moreover, deterministic
generation of large entangled states represents the most
remarkable attribute of CV fields. CV quantum optics has also
become a viable route for quantum computation where realistic
squeezing levels above 10 dB~\cite{FukuiPhysRevX8_2018} are
sufficient to achieve fault tolerance~\cite{MenicucciPRL112_2014}.
In addition, squeezed states have been found extraordinarily
useful in detection of the gravitational
waves~\cite{Schnabel2017}.

Although homodyne tomography \cite{Lvovsky2009} of Gaussian states
can be reduced to just 2 orthogonal cuts, the use of a local
oscillator requires advanced experimental setups. Nevertheless, at
present, there exists an alternative promising for practical
applications: photon-number-resolving detectors \cite{Mandel1995}.
The fact that they cannot capture the detected-state phase
properties does not have to be a drawback. This especially occurs
when complete characterization/identification of a measured state
does not require phase information due to its properties. Such
states can be recognized when we determine their quantum universal
invariants (QUIs) \cite{Sudak2025} and these QUIs are uniquely
determined using only photocount measurements. Such states then
immediately become extraordinarily promising for practical
applications in various CV-based metrology and quantum information
processing protocols.

Here, we show that symmetric three-beam Gaussian states (STBGSs)
belong to this kind of states. We note that these states are also
known as 3-mode squeezed thermal states and include, among others,
the noisy GHZ/W states \cite{Adesso2004, Adessopra73_2006}. They
have already demonstrated their effectiveness in the
implementation of several CV protocols \cite{PirandolaPRA68_2003,
Yonezawanature431_2004, FerraroPRA72_2005}. According to theory,
their full identification is reached when we know their 1-, 2-,
and 3-beam purities. The main result of our investigations - the
formula that gives the 3-beam purity in terms of intensity moments
up to the sixth order, together with the formula for the 2-beam
purity discovered recently \cite{BarasinskiPRL2023}, then allows
for this identification. Subsequently, different forms of quantum
correlations can be characterized and classified, including the
genuine 3-beam entanglement quantified by the Gaussian cotangle
\cite{Adesso_NJP2006}. Finally, using experimental 3-beam Gaussian
fields generated effectively with the help of parametric
down-conversion~\cite{PerinaJr.2021, BarasinskiPRL2023},
practicality of the developed approach is demonstrated.

The success in qualitative simplification of the experimental
identification of STBGSs poses a natural question: What are the
groups of states whose identification does not require the full
quantum state reconstruction and what parameters (QUIs) are necessary
for the state identification? This question gains particular significance
in multipartite systems, where the structure of quantum correlations
is more intricate and complex 
\cite{GiedkePPA64_2001} and thus potentially more attractive for various applications.
Moreover, our experimental results for 3-beam fields already provide certain insight into
understanding whether and how various types of quantum correlations are distributed over many different systems.

The paper is organized as follows. The structure and
parametrization of 3-beam symmetric Gaussian states are discussed
in Sec.~II. Their entanglement (steering) is theoretically
analyzed in Sec.~III (IV). Experimental characterization of STBGSs
using intensity moments of different
orders is discussed in Sec.~V. Also the experimental 1-, 2-, and
3-beam purities and seralian are given in Sec.~V, together with
other experimental characteristics including mean photon number,
the Fano factor, sub-shot-parameter, and the Kullback-Leibler
divergence. The properties of the observed genuine tripartite
entanglement are analyzed in Sec.~VI, together with those
belonging to the observed steering. The properties of the
corresponding GHZ/W states are discussed in Sec.~VII. Sec.~VIII
brings conclusions. Estimates and bounds for the quantities
characterizing STBGSs are discussed in
Appendix~A assuming knowledge of the intensity moments up to
fourth order. The role of the number of measurement repetitions in
the determination of experimental errors is addressed in
Appendix~B.

\section{Structure of three-beam symmetric Gaussian states}

We consider a CV 3-beam single-mode symmetric
system~\cite{comment} described in the Hibert space $\mathcal{H} =
\bigotimes_{j=1}^{3}\mathcal{H}_j$, where $\mathcal{H}_j$ stands
for the infinite-dimensional Fock space of beam $ j $. Let $\hat
x_j = (\hat a_j+\hat a_j^\dag)$ and $\hat p_j= -i(\hat a_j-\hat
a_j^\dag)$ be the quadrature operators with $a_j$ ($a_j^\dag$)
denoting the photon annihilation (creation) operator acting on
$\mathcal{H}_j$. The quadrature operators are then collected in
the vector $\hat X = (\hat x_1, \hat p_1,\hat x_2, \hat p_2,\hat
x_3, \hat p_3,)$ with the canonical commutation relations
expressed as $[\hat X_j,\hat X_k] = 2 i \Omega_{jk}$, where $
\boldsymbol{\Omega} = \bigoplus_{j=1}^{3}
    \begin{pmatrix}
  0 & 1\\
  -1 & 0
\end{pmatrix}$
is the symplectic form. Then, any Gaussian state with its
statistical operator $\hat{\rho}_3$ is fully specified, up to
local displacement, by its covariance matrix (CM)
$\boldsymbol{\sigma}_3$ with the matrix elements given by
$[\boldsymbol{\sigma}_3]_{jk} = \frac{1}{2} (\langle\hat X_j \hat X_k +\hat
X_k \hat X_j \rangle )  - \langle \hat X_j\rangle\langle\hat
X_k \rangle  $. The CM must fulfill the Robertson--Schr\"{o}dinger
uncertainty relation $\boldsymbol{\sigma}_3 + i
\boldsymbol{\Omega} \geq 0$ \cite{RobertsonPR34_1929,
Schrodinger1930, SerafiniPRL96_2006}.

For future convenience, let us write the CM $\boldsymbol{\sigma}_3$ of a STBGS in terms of two-by-two submatrices as
\begin{equation}\label{eq:CM3}    
    \boldsymbol{\sigma}_3 =  \begin{pmatrix}
\boldsymbol{\alpha} & \boldsymbol{\gamma}& \boldsymbol{\gamma} \\
\boldsymbol{\gamma} & \boldsymbol{\alpha} & \boldsymbol{\gamma} \\
\boldsymbol{\gamma} &\boldsymbol{\gamma} & \boldsymbol{\alpha}\\
\end{pmatrix},
\end{equation}
where $\boldsymbol{\alpha} = \text{diag}(a, a)$ is the local CM
corresponding to the reduced state of one beam and
$\boldsymbol{\gamma} = \text{diag}(c_+,c_-)$. We note that Eq.
\eqref{eq:CM3} refers to the so-called standard form
\cite{Simon2000, DuanPRL84_2000}. In our approach, we rely on an
alternative parameterization of $\boldsymbol{\sigma}_3$ that uses
three out of four state invariants \cite{Sudak2025}. They admit
direct interpretation for generic Gaussian states and include
global purity $\mu_3 =  \text{Det}(\boldsymbol{\sigma}_3)^{-1/2}$,
marginal k-beam purities $\mu_k =
\text{Det}(\boldsymbol{\sigma}_k)^{-1/2}$ for $k=1,2$, and
two-beam Seralian $\Delta_2 = 2\text{Det}(\boldsymbol{\alpha}) +
2\text{Det}(\boldsymbol{\gamma})$.

By selecting $\mu_{1}$, $\mu_2$, and $\Delta_2$ as independent parameters (motivated by Ref.~\cite{SerafiniPRA71_2005}), the following relationship can be introduced
\begin{eqnarray}   
    a &=& \frac{1}{\mu_1}, \hspace{5mm} 
    c_{\pm} = \frac{1}{4}\sqrt{\mu_1^2 \left(\Delta_2^2-\frac{4}{\mu_2^2}\right)}\pm\epsilon, \nonumber\\
    & & \epsilon \equiv \frac{1}{4}\sqrt{\frac{(4\mu_1^2-\mu_1^4\Delta_2)^2}{\mu_1^6}-\frac{4\mu_1^2}{\mu_2^2}}.
    \label{eq:p1}
\end{eqnarray}
From Eq. \eqref{eq:p1} one immediately finds that the necessary
conditions for a physical state, $c_{\pm} \in \mathbb{R}$, and the
Robertson--Schr\"{o}dinger uncertainty relation, are
simultaneously expressed as
\begin{eqnarray}
   & \mu_1^2\le\mu_2\le\mu_1 \hspace{15mm}  {\rm and} &
    \label{eq:domain}    \\
   & \mathcal{F}(\mu_1,\mu_2) \le \Delta_2 \le
     \min \Big\{\frac{4}{\mu_1^2}-\frac{2}{\mu_2},1+\frac{1}{\mu_2^2}\Big\}, &
     \label{eq:delta3} \\
  & \mathcal{F}(\mu_1,\mu_2) = \left\{ \begin{array}{ll}
  \dfrac{2}{\mu_2} & \textrm{for~} \mu_1^2 \leq \mu_2 \leq \dfrac{4 \mu_1^2}{3 + \mu_1^2}\\
  & \\
  \dfrac{1}{3}\left( 2 + \frac{6}{\mu_1^2}-\omega\right) & \textrm{for~} \dfrac{4 \mu_1^2}{3 + \mu_1^2}<\mu_2\le \mu_1
\end{array} \right. , & \nonumber
\end{eqnarray}
where $\omega = \sqrt{\frac{(\mu_1^2+3)^2}{\mu_1^4} - \frac{12}{\mu_2^2}}$.

Importantly, both purities $\mu_1$ and $\mu_2$ are experimentally
available \cite{BarasinskiPRL2023} using the measured intensity
moments that are the normally ordered photon-number moments. They
are obtained by the Stirling numbers from the measured
photon-number moments \cite{Perina1991}. For the Seralian $
\Delta_2 $, however, only its upper and lower bounds are
derived~\cite{Sudak2025}, independently of Eq. \eqref{eq:delta3}.
Nevertheless, in the analyzed symmetric case, $\Delta_2$ can readily be
determined from purity $\mu_{3}$, as the above analyzed parameters must satisfy the relation 
\begin{eqnarray}
\mu_3^{-2} &=&  \text{Det}(\boldsymbol{\sigma}_3) =\frac{1}{2 \mu_1^2 \mu_2^3} \bigg(6 \mu_2 - 3 \mu_2^3 \Delta_2^2 + \mu_1^2 \mu_2^3 \Delta_2^3 \nonumber\\
&+& \sqrt{\mu_2^2 \Delta_2^2-4} \Big(3 \mu_2^2 \Delta_2 -2 \mu_1^2 - \mu_1^2 \mu_2^2 \Delta_2^2\Big)\bigg).
\label{eq:purity3}
\end{eqnarray}

This implies that the purities of 2- and 3-beam systems, along
with the single-beam purity, are sufficient to fully identify and
thus characterize a STBGS. However, the
experimental determination of the 3-beam purity using intensity moments
is prone to significant errors. This is so since the determination of
$\mu_3$ requires intensity moments up to sixth order
\cite{Sudak2025}. Therefore, following the approach in Ref.
\cite{BarasinskiPRL2023}, it is advisable to parameterize the
STBGSs by Eq.~\eqref{eq:p1} and replace
the Seralian $ \Delta_2 $ by its minimum and maximum allowed
values. In Ref. \cite{BarasinskiPRL2023}, this method proved
useful for obtaining the negativity of 2-beam fields with higher
degrees of entanglement. Moreover, combining
Eq.~\eqref{eq:purity3} with Eq.~\eqref{eq:delta3} we reveal the upper
and lower bounds for $\mu_3$ in terms of the marginal purities
$\mu_1$ and $\mu_2$.

\begin{table}
 \caption{Conditions for observation of different types of correlations characterizing entanglement and Gaussian steering $A\rightarrow BC$ and $BC\rightarrow A$
 in the system with three beams $A$, $B$, and $C$ expressed in terms of purities $ \mu_1 $ and $ \mu_2 $.}
\begin{ruledtabular}
        \begin{tabular}{cc}
Degrees of purity & Correlations \\ \hline
    \multicolumn{2}{c}{Entanglement $A - BC$} \\ \hline
$\mu_1^2\le \mu_2 \le \frac{3 \mu_1^2+\sqrt{9+62 \mu_1^2-7 \mu_1^4}-3}{10 - 2 \mu_1^2}$ &  coexistence region I\\
$\frac{3 \mu_1^2+\sqrt{9+62 \mu_1^2-7 \mu_1^4}-3}{10 - 2 \mu_1^2} < \mu_2 \le \frac{\mu_1}{\sqrt{2 - \mu_1^2}}$ &  coexistence region II\\
$\frac{\mu_1}{\sqrt{2 - \mu_1^2}} < \mu_2 \le \mu_1$ & entangled states\\ \hline
    \multicolumn{2}{c}{Gaussian steering $A\rightarrow BC$} \\ \hline
$\mu_1^2\le \mu_2 \le \frac{\sqrt{16 \mu_1^2+9}-3}{2}$ & unsteerable states \\
$\frac{\sqrt{16 \mu_1^2+9}-3}{2} < \mu_2 \le \frac{\sqrt{3} \mu_1}{\sqrt{4-\mu_1^2}}$ & coexistence region\\
$\frac{\sqrt{3} \mu_1}{\sqrt{4-\mu_1^2}} < \mu_2 \le\mu_1$ & steerable states\\  \hline
    \multicolumn{2}{c}{Gaussian steering $BC \rightarrow A$} \\ \hline
$\mu_1^2\le \mu_2 \le \frac{4\mu_1^2}{3+\mu_1^2}$ & unsteerable states \\
$\frac{4\mu_1^2}{3+\mu_1^2} < \mu_2 \leq \frac{\sqrt{2} \mu_1}{\sqrt{3-\mu_1^2}}$ & coexistence region\\
$\frac{\sqrt{2} \mu_1}{\sqrt{3-\mu_1^2}} < \mu_2\le\mu_1$ & steerable states\\
        \end{tabular}
    \end{ruledtabular}

 \label{tab_1}
\end{table}

Despite their simple form, the 3-beam symmetric states $\sigma_3$
exhibit a remarkably rich structure of nonclassical correlations.
Whereas their entanglement is addressed in detail in the following
Sec.~III, their steering properties are deeply analyzed in
Sec.~IV.

\section{Entanglement in Symmetric Three-Beam Gaussian states}
\label{Entanglement}

Positivity of a partially transposed CM,
$\tilde{\sigma}$, has been established as the necessary and
sufficient condition for determining the separability of (1 + N)-beam
and bisymmetric (M + N)-beam Gaussian states
\cite{Simon2000,DuanPRL84_2000,SerafiniPRA71_2005}. This so-called
positive partial transpose (PPT) criterion provides a clear qualitative
characterization of the entanglement present in such states.
Specifically, a Gaussian state described by the CM $\sigma$ is separable
if and only if all symplectic eigenvalues $\tilde{v}_i$ of its partially transposed CM
$\tilde{\sigma}$ satisfy $\tilde{v}_i \ge 1$. Though the converse does not necessarily
hold in a generic case, as bound entangled states may also meet the PPT criterion.

For tripartite Gaussian states, a comprehensive qualitative classification of
entanglement was presented in Ref.~\cite{GiedkePPA64_2001}, which distinguishes
five classes according to their separability under different bipartitions. Among
these, only three are relevant for STBGSs: fully inseparable (class~1), bound
entangled (class~4), and fully separable (class~5) states.
Based on the single-mode and two-mode purities, $\mu_1$ and $\mu_2$, the corresponding
classification is summarized in Table~\ref{tab_1}.
Notably, only class 1 states (though not all) can be uniquely
identified using marginal purities. For the remaining cases, two
coexistence regions emerge: Region I, containing states from
classes 4 and 5 (depending on the value of $\Delta_2$), and Region II,
comprising states from classes 1, 4, and 5. The domain of tripartite
entanglement shown in Table~\ref{tab_1} coincides with that of pairwise
entanglement~\cite{Adesso2004b}, reflecting the promiscuous sharing of CV
entanglement \cite{Adesso_NJP2006}. However, a notable discrepancy
arises in coexistence Region II, where states exhibiting tripartite entanglement
without any bipartite entanglement are found (see below).

These results emerge in the following calculations. Without loss of generality,
we consider the bipartition of the CM $\sigma_3$ in the form $A-BC$, where
$A$, $B$, and $C$ denote the field beams. In this case, the symplectic eigenvalues of the
partially transposed CM $\tilde{\sigma} = \Lambda\sigma\Lambda$,
where $\Lambda = \text{diag}(1,-1,1,1,1,1)$, are given by
\begin{eqnarray}
\tilde{v}_1^2 &=& (a-c_-)(a-c_+) , \nonumber\\
\tilde{v}_{\pm}^2 &=& \frac{1}{2} \bigg(2a^2-3c_-c_+ + a(c_-+c_+)
\pm \theta \bigg) \label{PPT_eigenvalues} ,
\end{eqnarray}
where
\begin{displaymath}
    \theta = \sqrt{9 a^2 c_-^2 + 2 a c_- c_+(c_--7a)+(9a-7c_-)(a+c_-)c_+^2}.
\end{displaymath}

Utilizing Eq.~(\ref{eq:p1}), it becomes apparent that
$\tilde{v}_1$ and $\tilde{v}_{+}$ remain greater than or equal to
$1$ throughout the entire domain delineated by
Eqs.~(\ref{eq:domain}) and (\ref{eq:delta3}). Consequently, the
entanglement characterization of $\sigma_3$ is solely based on
$\tilde{v}_{-}$. Through this analysis, we have determined that
the state $\sigma_3$ satisfies the PPT criterion ($\tilde{v}_{-}
\ge 1$) when
\begin{eqnarray}
\label{eq:full_sep} \mu_1^2\le \mu_2 \le \dfrac{3
\mu_1^2+\sqrt{9+62 \mu_1^2-7 \mu_1^4}-3}{10 - 2 \mu_1^2},
\end{eqnarray}
for $0<\mu_1\le 1$ and all $\Delta_2$ values specified in
Eq.~(\ref{eq:delta3}). This parameter range is hereafter referred
to as Region I. Our analysis indicates, however, that satisfying
the PPT criterion within Region I does not necessarily guarantee
the separability of the states $\sigma_3$. To illustrate this, let
us consider a specific example of $\sigma_3$, namely the noisy
GHZ/W state. These states are fully characterized by their 1- and
2-beam purities $ \mu_1 $ and $ \mu_2$ or equivalently by the
parameters $a=\mu_1^{-1}$ and $n=\mu_1/\mu_2$ (cf. Ref.
\cite{AdessoNPJ2007}).
We examine two noisy GHZ/W states within Region I:
$\{\mu_1, \mu_2\} = \{0.5, 0.25\}$ and $\{\mu_1,
\mu_2\} = \{0.8, 0.652352\}$. According to the classification of
Refs.\cite{CHEN2005121, AdessoNPJ2007}, the former corresponds to
a separable state (class~5), while the latter represents a bound
entangled state (class~4). Thus, Region~I contains both classes~4
and~5, with the specific class determined by $\Delta_2$.

\begin{figure*}[htp]  
\centering
\begin{tabular}{c} 
{\includegraphics[width=0.35\linewidth]{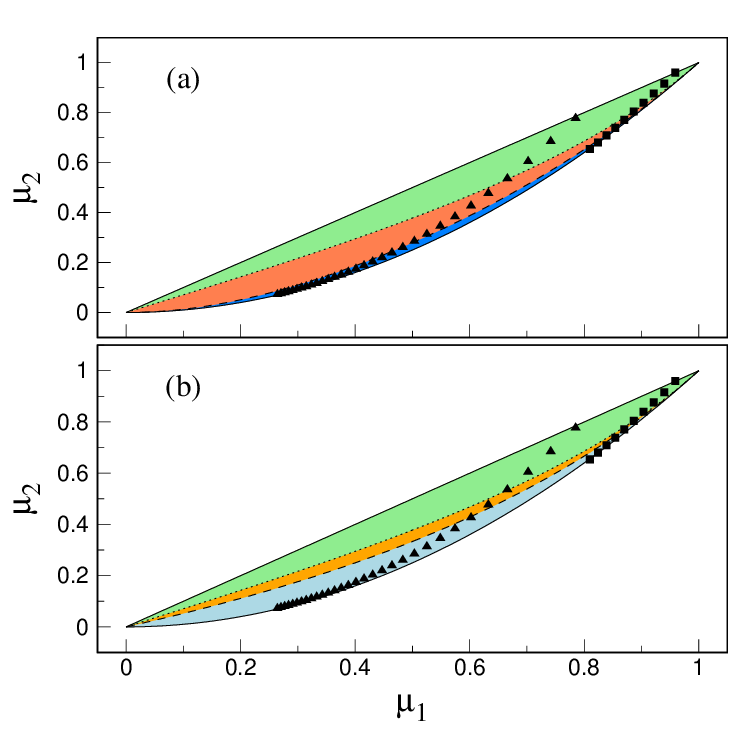}}
\end{tabular}~
\begin{tabular}{c} 
{\includegraphics[width=0.3\linewidth]{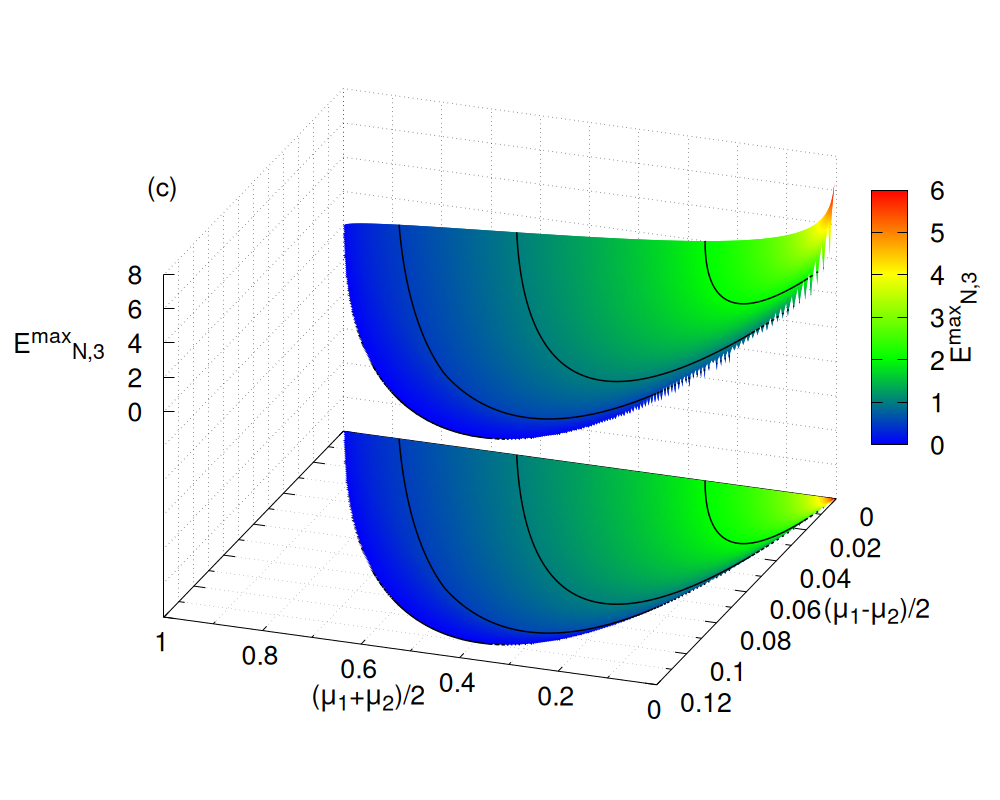}}
\\
{\includegraphics[width=0.3\linewidth]{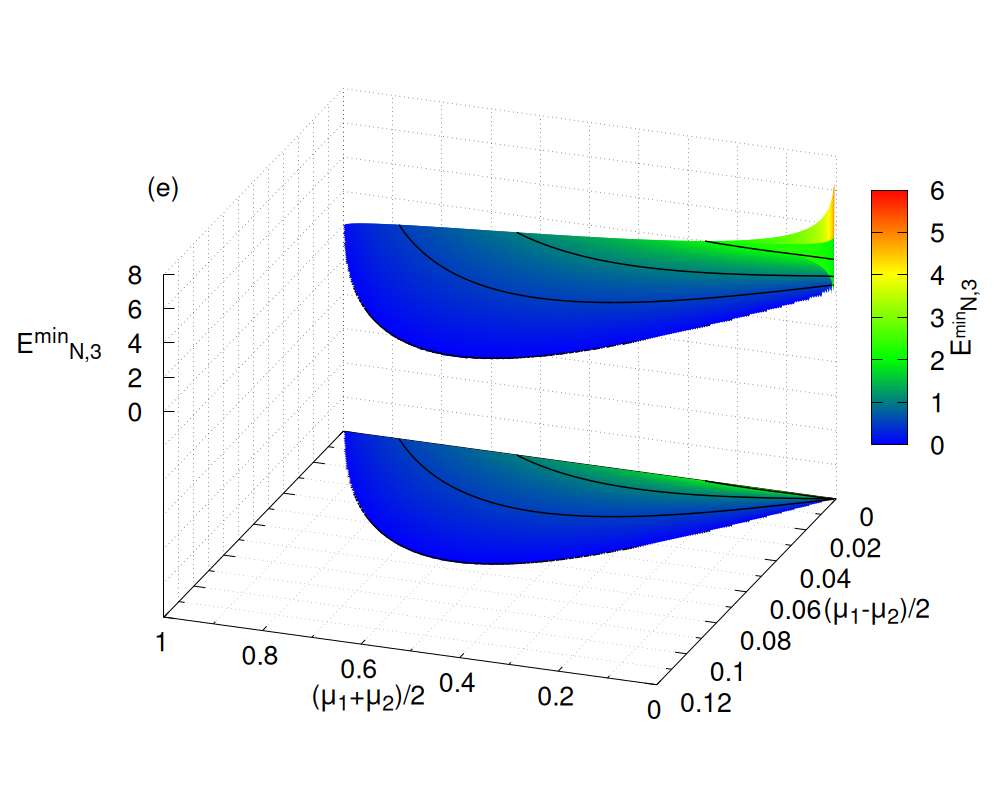}}
\end{tabular}~
\begin{tabular}{c} 
{\includegraphics[width=0.3\linewidth]{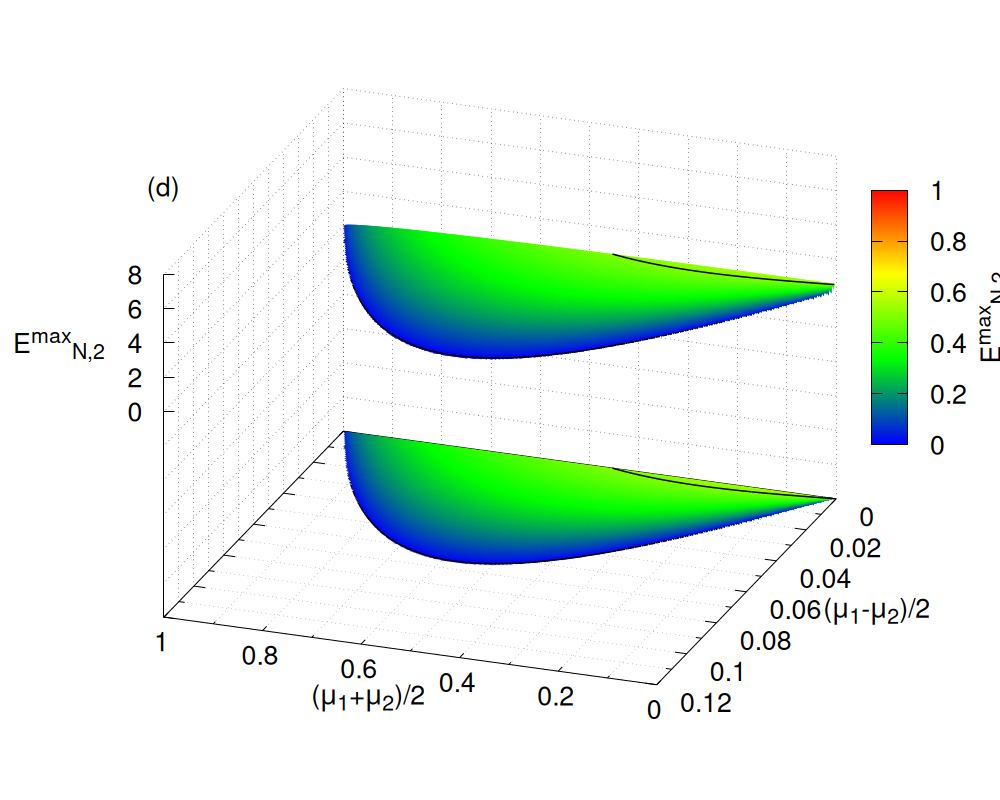}}
\\
{\includegraphics[width=0.3\linewidth]{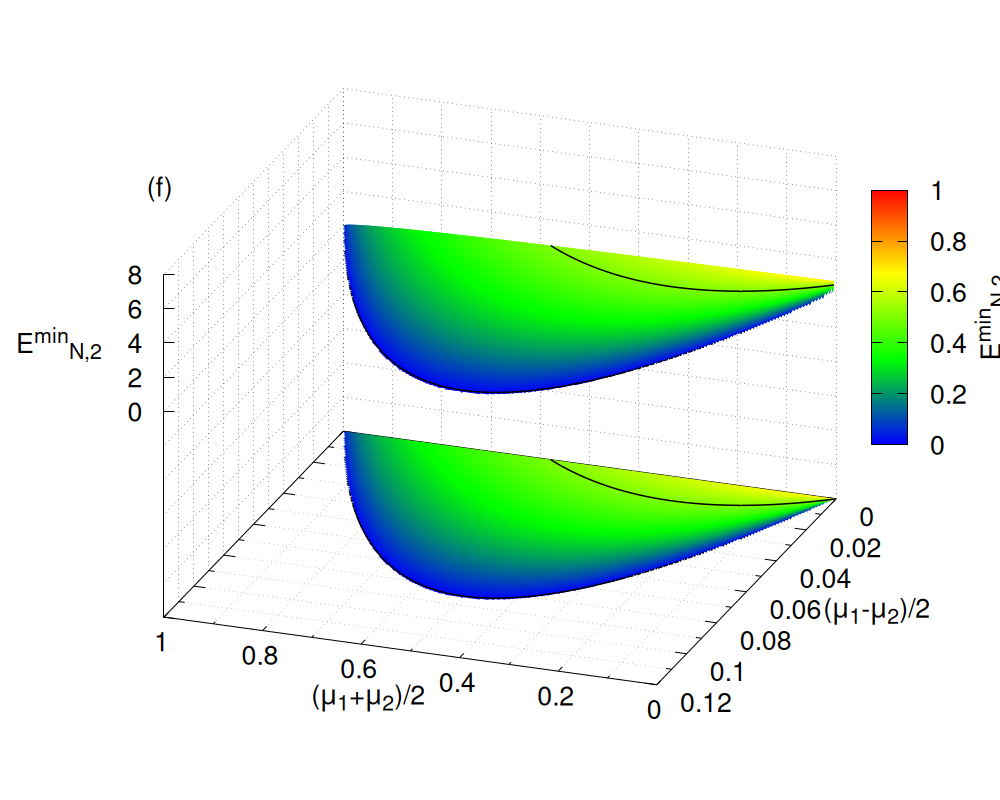}}
\end{tabular}
 \caption{Two- and three-beam entanglement analyzed in the plane
 spanned by 1- and 2-beam purities $ \mu_1 $ and $ \mu_2 $. Panel
 (a) highlights areas differing in 3-beam entanglement properties:
 Region I (coexistence of separable and bound entangled states) is
 shown in blue, Region II (coexistence of separable, bound
 entangled, and fully entangled states) is depicted in orange, the
 fully entangled region is drawn in green. Panel (b) focuses on
 2-beam entanglement: area with 2-beam separable (entangled) states
 is shown in light blue (green). In the light-orange area separable
 and entangled states coexist. These regions are classified using
 the criteria of Ref.~\cite{Adesso2004b}. Isolated black boxes (black
 triangles) refer to the experimental data discussed
 for $ M = 6.7 $ ($ M = 40 $) independent modes
 Upper and lower bounds on three-beam (two-beam) logarithmic negativity $E_{N,3}$ [$E_{N,2}$]
 are shown in panels (c) [(d)] and (e) [(f)], respectively, as they depend on
 'rotated' marginal purities $(\mu_1 + \mu_2)/2$ and $(\mu_1 - \mu_2)/2$.
  In panels (c) and (e), the black solid curves indicate contour
 levels of $E_{N,3}$ equal to 0.5, 1, and 2. In panels (d) and (f),
 the black solid curves correspond to $E_{N,2} = 0.5$. }
\label{fig1}
\end{figure*}

On the other hand, one can find that $\tilde{v}_{-} <1$, i.e. the
CMs $\sigma_3$ describe fully inseparable states (class 1), if
\begin{equation}
    \frac{\mu_1}{\sqrt{2 - \mu_1^2}}\le\mu_2\le\mu_1,
\end{equation}
where $0<\mu_1\le 1$ and all $\Delta_2$ values specified in
Eq.~(\ref{eq:delta3}). Between these two regions (see Fig.
\ref{fig1}), we find the states that belong to one of three
classes: class 1, class 4, or class 5. For example, a separable
state in this region is represented by the noisy GHZ/W state with
parameters $\{\mu_1, \mu_2\} = \{0.5, 0.28\}$, while a bound
entangled state in this region is exemplified by the noisy GHZ/W
state with parameters $\{\mu_1, \mu_2\} = \{0.5, 1/(2\sqrt{3})\}$.

In addition to the state classification outlined above, it is also
feasible to determine the lower and upper bounds of the 2-beam and
3-beam logarithmic negativities $ E_{N,2} $ and $ E_{N,3}$
\cite{Hill1997,Horodecki2009,Adesso2007}. They are defined in
general as $E_{N} \equiv \ln ||\tilde{\rho}||_1$, where
$||\cdot||_1$ denotes the trace norm. For Gaussian states, this
measure can be determined in terms of the symplectic spectrum
$\tilde{v}$ of the partially transposed CM $\tilde{\sigma}$,
$E_{\mathcal{N}} = \max \{0,-\sum_{i:\tilde{v}_i<1}  \ln
\tilde{v}_i\}$. In this case, only one eigenvalue can take values
smaller than 1, and the logarithmic negativity is defined as
$E_{N,3} = \max \{0,- \ln \tilde{v}_-\}$. Moreover, since
$\tilde{v}_-$ in Eq.~(\ref{PPT_eigenvalues}) is monotonic with
respect to $\Delta_2$, the lower and upper bounds of logarithmic
negativity are determined by the extremal values of $\Delta_2$ for
the fixed 1- and 2-beam purities $ \mu_1 $ and $ \mu_2 $.
Specifically, the bounds are given by $E_{N,3}^{\rm min} = E_{N}
(\max\{\Delta_2\})$ and $E_{N,3}^{\rm max} = E_{N}
(\min\{\Delta_2\})$. Given the complexity of the resulting
expressions, we do not present them explicitly. Instead, we draw
$E_{N,3}^{\rm min}$ and $E_{N,3}^{\rm max} $ for the symmetric
three-beam Gaussian states as they depend on the 'rotated'
marginal purities $ (\mu_1+\mu_2)/2 $ and $ (\mu_1-\mu_2)/2 $ in
Figs.~\ref{fig1}(c-f).

Finally, the bounds on the logarithmic negativity can be employed
to characterize genuine tripartite entanglement through the
residual cotangle \cite{Adesso_NJP2006}, defined as
\begin{equation}
  T_\pi  =  E_{N,3}^2 - 2 E_{N,2}^2,
 \label{cotangle}
\end{equation}
where $E_{N,2} = -\frac{1}{2} \ln \left(a^2 - a |c_m -
c_p| - c_m c_p\right)$. Using Eq.~(\ref{eq:p1}) and following a
series of lengthy but straightforward
calculations, one can demonstrate that the residual cotangle
$T_\pi $ is a monotonic function of $\Delta_2$. Consequently, it
follows that $ T_\pi $ attains its maximum and minimum values at
the respective boundary values of $\Delta_2$.

\section{Steering in Symmetric Three-Beam Gaussian states}
\label{Steering}

Gaussian steering is a stronger form of quantum correlation than entanglement~\cite{Kogias2015}, playing a key role in asymmetric quantum information tasks such as one-sided device-independent protocols. Due to its directional nature, it exhibits a rich and structured hierarchy of correlations that depend on state purity and symmetry. A detailed classification of the Gaussian-steering properties of STBGSs, based on their marginal purities, is presented in Table~\ref{tab_1}.

Interestingly, STBGSs show no Gaussian steering between individual beams. To demonstrate this, recall that the measure of Gaussian steerability from $M$ to $1-$beam subsystems simplifies to~\cite{Kogias2015}
\begin{eqnarray}
\mathcal{G}^{M \rightarrow 1}(\mathbf{\sigma}) = \max\left\{0,\ln{\frac{\mu_{M+1}}{\mu_{1}}}\right\},
\label{eq:Gaussian_stearingP}
\end{eqnarray}
where $\mu_1$ denotes the purity of a single beam and $\mu_{M+1}$ the purity of the joint $(M + 1)-$beam subsystem. From Eq.~\eqref{eq:Gaussian_stearingP}, it follows that fully symmetric states $\sigma_3$ cannot exhibit Gaussian steering of type $1 \rightarrow 1$, as the condition $\mu_2 > \mu_1$ violates the constraint given in Eq.~(\ref{eq:domain}).

\begin{figure*}[htp]
\centering
\begin{tabular}{c}    
{\includegraphics[width=0.35\linewidth]{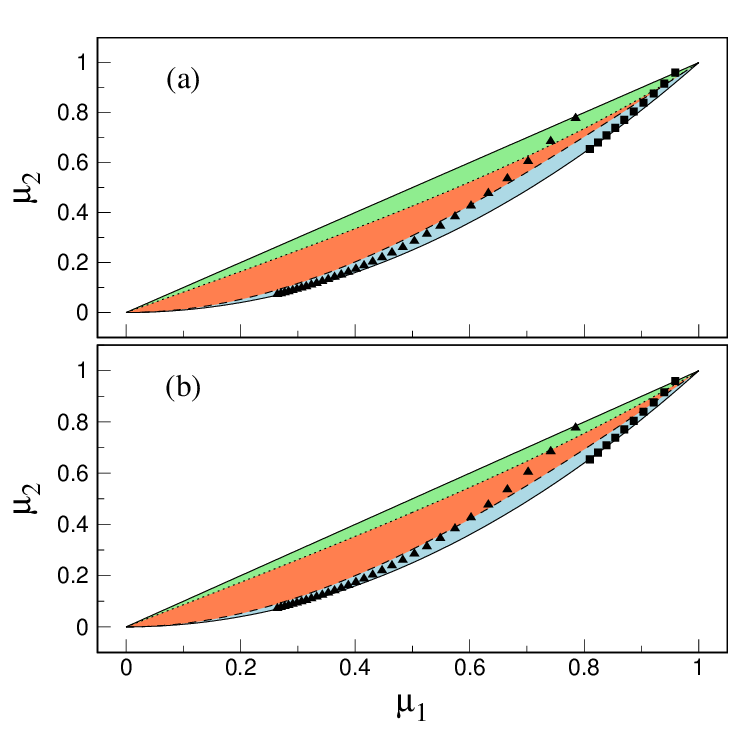}}
\end{tabular}~
\begin{tabular}{c} 
{\includegraphics[width=0.3\linewidth]{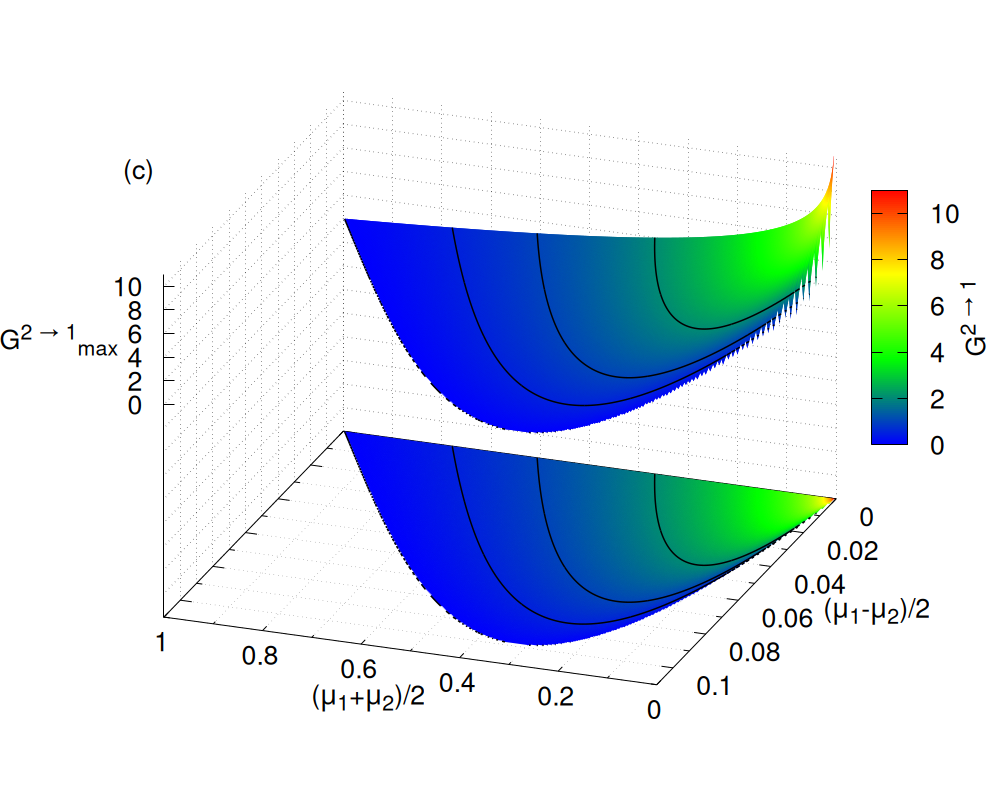}}
\\
{\includegraphics[width=0.3\linewidth]{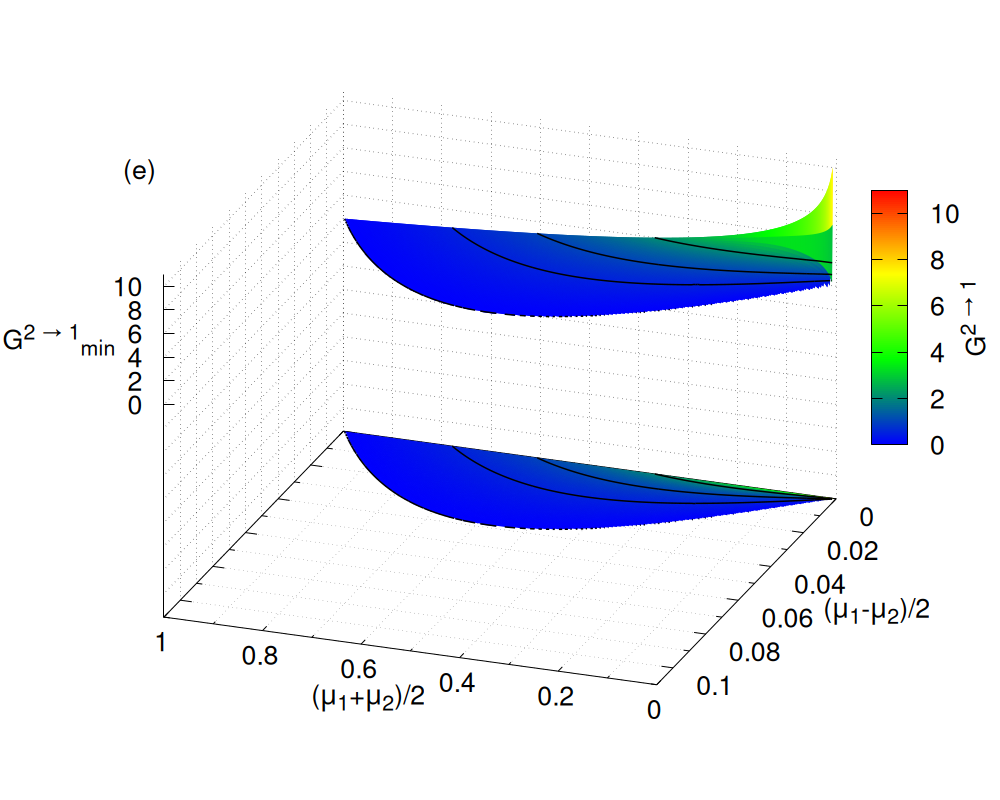}}
\end{tabular}~
\begin{tabular}{c}    
{\includegraphics[width=0.3\linewidth]{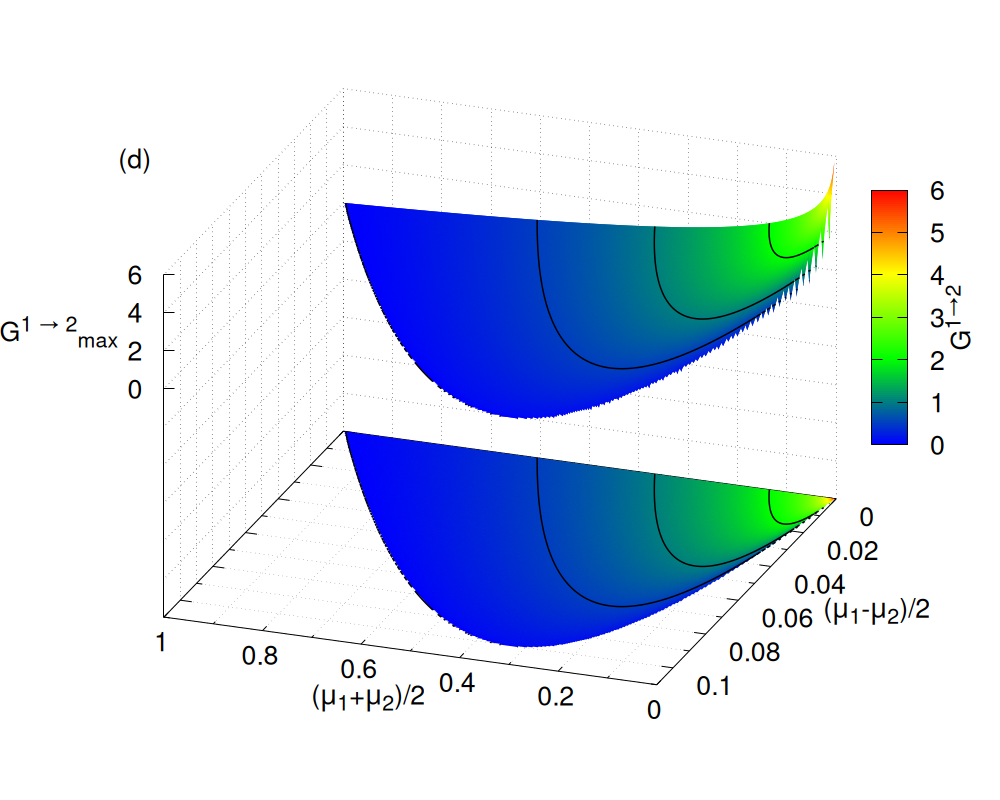}}
\\
{\includegraphics[width=0.3\linewidth]{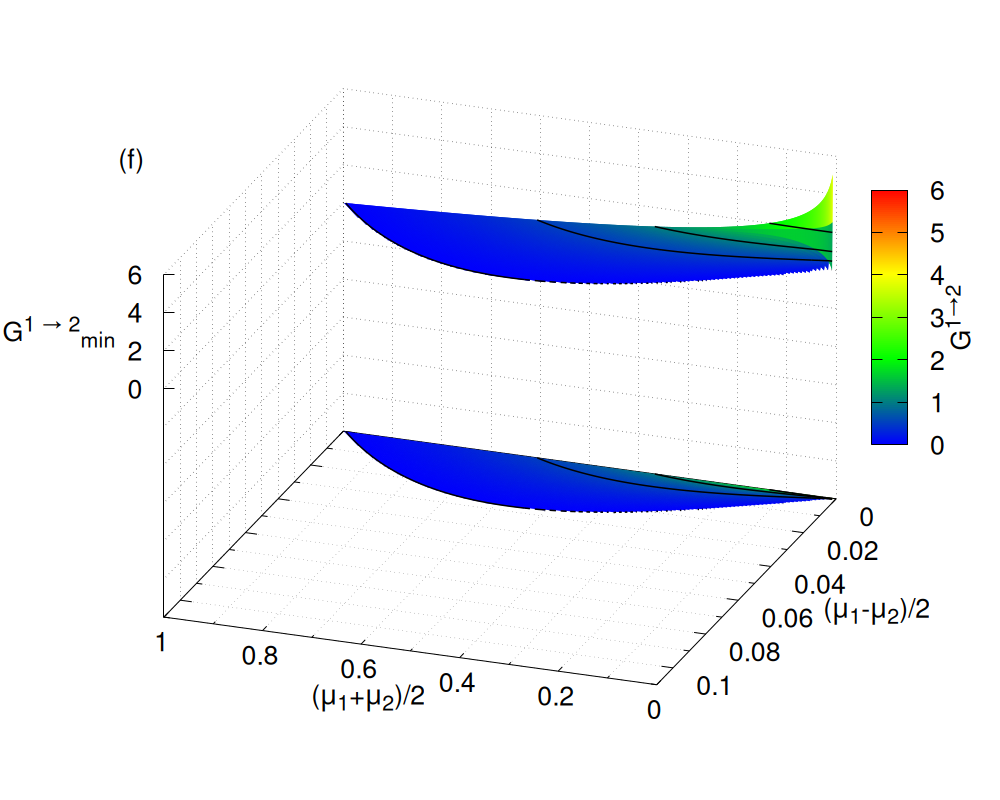}}
\end{tabular}
 \caption{Three-mode Gaussian steering: Steerable region,
 coexistence region, and unsteerable region for steering parameters
 (a) $ {\cal G}^{2\rightarrow 1}$ and (b) $ {\cal G}^{1\rightarrow
 2}$ are drawn in turn in green, orange, and blue color in the
 plane spanned by 1- and 2-beam purities $ \mu_1 $ and $ \mu_2 $.
 Isolated black boxes (black triangles) refer to the experimental data for $ M = 6.7 $ ($ M = 40 $).
 Upper and lower bounds $ {\cal G}^{2\rightarrow
 1}_{\rm max} $ [$ {\cal G}^{1\rightarrow 2}_{\rm max} $] and $ {\cal G}^{2\rightarrow
 1}_{\rm min} $ [$ {\cal G}^{1\rightarrow 2}_{\rm min} $]  on the steering parameter
 as they depend on 'rotated' marginal purities $(\mu_1+\mu_2)/2 $ and $ (\mu_1-\mu_2)/2 $
 are drawn in (c) [(d)] and (e) [(f)], respectively. In (c) -- (f), black solid curves correspond to $ {\cal G}^{2\rightarrow 1}_{\rm max} $, $ {\cal G}^{1\rightarrow 2}_{\rm max} $, $ {\cal G}^{2\rightarrow 1}_{\rm min} $, and $ {\cal G}^{1\rightarrow 2}_{\rm min} $ equal to 0.5, 1, and 2. }
\label{fig2}
\end{figure*}

However, Gaussian steering emerges between the two-beam and single-beam subsystems. In the $2 \rightarrow 1$ scenario, the ratio of purities is given by
\begin{eqnarray}   
\frac{\mu_{3}}{\mu_{1}} = \frac{\mu_1^2 \mu_2 (\mu_2 \Delta_2 + \gamma)}{3 \mu_2 (-\mu_2 \Delta_2 + \gamma) + \mu_1^2 (4 + \mu_2 \Delta_2 (\mu_2 \Delta_2 - \gamma))},
\label{eq:G21}
\end{eqnarray}
where $\gamma = \sqrt{\mu_2^2\Delta^2_2 - 4}$. Equation~\eqref{eq:G21} is monotonic in $\Delta_2$, so its extrema occur at the boundaries of the allowed $\Delta_2$ domain. Consequently, $\mathcal{G}^{2\rightarrow 1}(\mathbf{\sigma}_3) > 0$ for
\begin{equation}
\frac{\sqrt{2}\mu_1}{\sqrt{3 - \mu_1^2}} \le \mu_2 \le \mu_1,
\label{eq:steering_regime}
\end{equation}
where $0 < \mu_1 < 1$ and for all attainable $\Delta_2$ values. The coexistence region, where steering occurs only for certain $\Delta_2$, is
\begin{equation}
\frac{4\mu_1^2}{3+\mu_1^2} \le \mu_2 < \frac{\sqrt{2}\mu_1}{\sqrt{3 - \mu_1^2}},
\end{equation}
as illustrated in Fig.~\ref{fig2}(a).

In the opposite configuration, $1 \rightarrow 2$, the measure of Gaussian steerability is defined as~\cite{Kogias2015}
\begin{equation}
\mathcal{G}^{1 \rightarrow 2}(\mathbf{\sigma_3}) = \max\left\{0,-\sum_{j:\bar \nu_j<1}\ln{\bar \nu_j}\right\},
\label{eq:Gaussian_stearing}
\end{equation}
where $\bar{\nu}_j$ are the symplectic eigenvalues of the Schur complement $M_{\sigma}$ to single beam CM~\cite{comment2}. For $\sigma_3$, the relevant symplectic eigenvalue is
\begin{equation}
\bar{\nu}^2 = \frac{3 \mu_2 (-\mu_2 \Delta_2 + \gamma) + \mu_1^2 (4 + \mu_2 \Delta_2 (\mu_2 \Delta_2 - \gamma))}{2 \mu_2^2},
\end{equation}
with $\gamma $ defined below Eq.~\eqref{eq:G21}. Since $\bar{\nu}$ is also monotonic in $\Delta_2$, one finds $\mathcal{G}^{1\rightarrow 2}(\mathbf{\sigma}_{3}) > 0$ for
\begin{equation}
\frac{\sqrt{3}\mu_1}{\sqrt{4 - \mu_1^2}} \le \mu_2 \le \mu_1,
\end{equation}
and all allowed $\Delta_2$. The coexistence region, where steering occurs only for specific $\Delta_2$, is
\begin{equation}   
\frac{\sqrt{16\mu_1^2 + 9} - 3}{2} \le \mu_2 < \frac{\sqrt{3}\mu_1}{\sqrt{4 - \mu_1^2}},
\end{equation}
as shown in Fig.~\ref{fig2}(b).

 Both measures, $\mathcal{G}^{2\rightarrow 1}(\mathbf{\sigma}_3)$ and $\mathcal{G}^{1\rightarrow 2}(\mathbf{\sigma}_3)$, are monotonic in $\Delta_2$ and thus reach their bounds for $\max\{\Delta_2\}$ and $\min\{\Delta_2\}$, respectively. The corresponding lower and upper bounds of ${\cal G}^{2\rightarrow 1}$ and ${\cal G}^{1\rightarrow 2}$ are plotted in Figs.~\ref{fig2}(c---f) as functions of the “rotated” marginal purities $(\mu_1 + \mu_2)/2$ and $(\mu_1 - \mu_2)/2$.

\section{Experimental investigation of quantum correlations in
three-beam symmetric Gaussian states}

The above theoretical characterization enables experimentally
reliable estimation of CV nonclassical correlations using
intensity moments up to fourth order through determination of
marginal purities, offering robust sufficient conditions and
analytical bounds. Furthermore, since CV logarithmic negativity
and Gaussian steering measures are monotonic functions of
$\Delta_2$, the experimental marginal purities also facilitate a
reliable estimation of entanglement. This result further enables
the estimation of genuine tripartite entanglement via the residual
cotangle defined in Eq.~(\ref{cotangle}). Accurate classification
of entanglement in the coexistence regions requires the knowledge
of $\Delta_2$, which can be obtained either directly from $\mu_3$
using intensity moments up to sixth order or estimated from the
experimental data (see Ref.\cite{Sudak2025})

\begin{figure}  
  \centerline{\includegraphics[width=0.85\hsize]{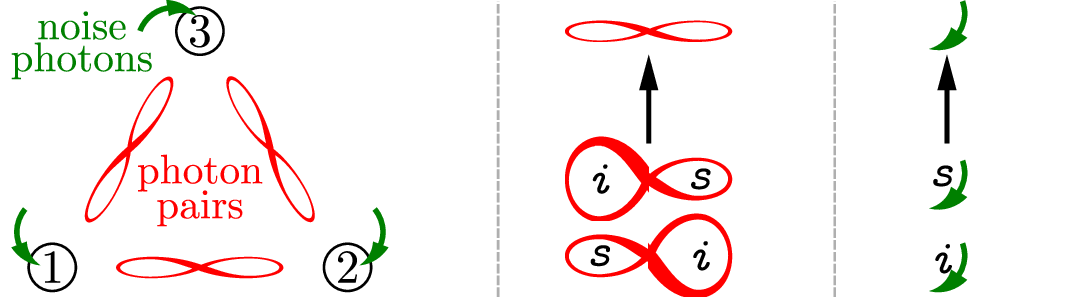}}
 \caption{Structure of 3-beam Gaussian states containing photon-pair (red $ \infty $)
and single-photon (green $ \rightarrow $) fields. Both fields can
be symmetrized by alternating the signal (s) and the idler (i)
beams (double $ \infty $ and $ \rightarrow $)} \label{fig3}
\end{figure}

Using the experimental photocount technique, a group of STBGSs
including those with genuine 3-beam entanglement was prepared and
investigated. The state preparation is based on the fact that the
Gaussian states are fully characterized just by their first- and
second-order amplitude moments and so they can in principle be
constructed from pairwise correlations embedded in twin beams. The
underlying principal scheme drawn in Fig.~\ref{fig3} uses three
twin beams whose constituting beams contribute differently to
beams 1, 2, and 3. Additional beams with increasing intensities
then form the noisy parts of STBGSs. Using the technique of
compound beams suggested in Ref.~\cite{PerinaJr.2021}, all needed
beams are derived from a sequence of measurements of photocount
distributions of identical weak twin beams by two single-photon
avalanche photodiodes (APDs). The experimental data characterizing
the analyzed STBGSs are then obtained by suitable concatenating
the data representing the signal and the idler photocounts of
these weak twin beams (for details, see Sec.~I and Fig.~1 in~Supplemental Material
(SM) \cite{SM}). We
note that, when constructing the states, we alternate the signal
and the idler beams to arrive at the Gaussian states symmetric in
both intensities and correlations. Whereas the correlated parts of
all analyzed states are composed of 12 weak twin beams, up-to 180
weak twin beams are used to form the noise parts of the analyzed
states.

The experimental data were transformed into photocount histograms
$ f(c_1,c_2,c_3) $ giving the probability of measuring
simultaneously $ (c_1,c_2,c_3) $ photocounts in beams $ (1,2,3) $.
Knowing the parameters of the detectors used, the corresponding
photon-number distribution $ p(n_1,n_2,n_3) $ was reconstructed
using the maximum-likelihood approach
\cite{Dempster1977,Vardi1993}. For the model accompanying the
experimental results (see Sec.~II in SM~\cite{SM}), parameters of weak twin
beams were determined (see Ref.~\cite{PerinaJr.2021}).

\begin{figure}  
  \centerline{\includegraphics[width=0.47\hsize]{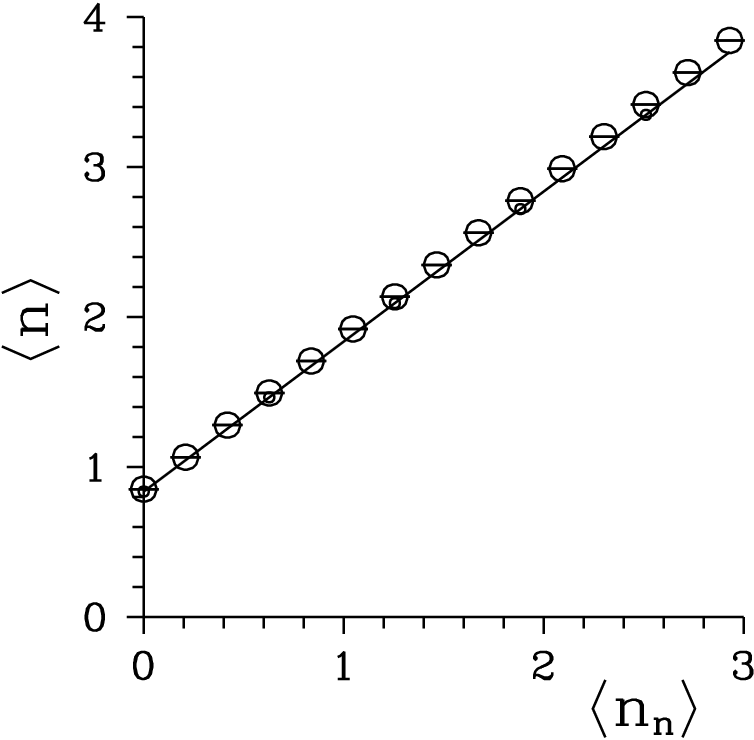}
    \hspace{2mm}
     \includegraphics[width=0.47\hsize]{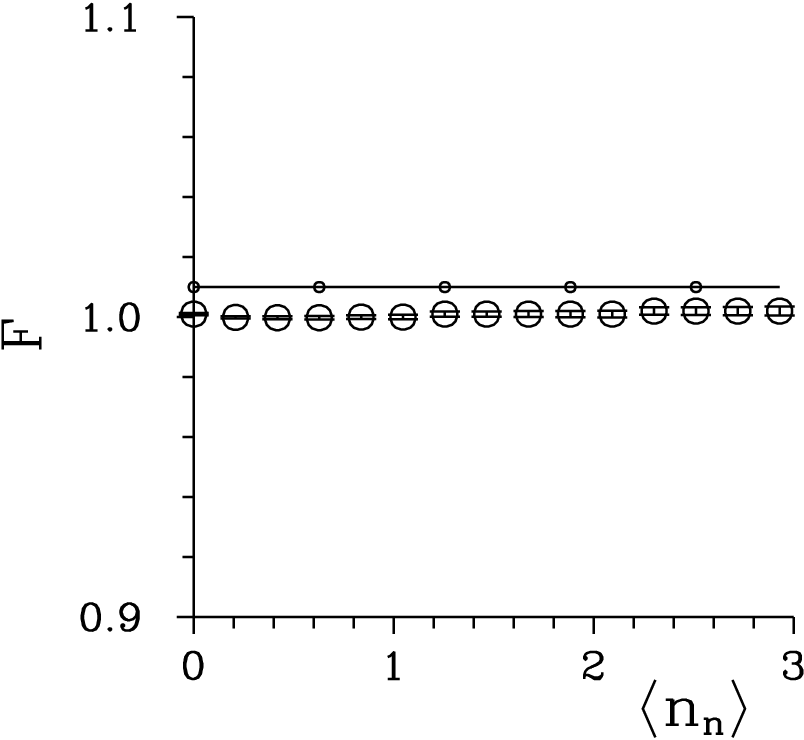}}
  \centerline{(a)\hspace{0.45\hsize} (b)}
  \centerline{\includegraphics[width=0.47\hsize]{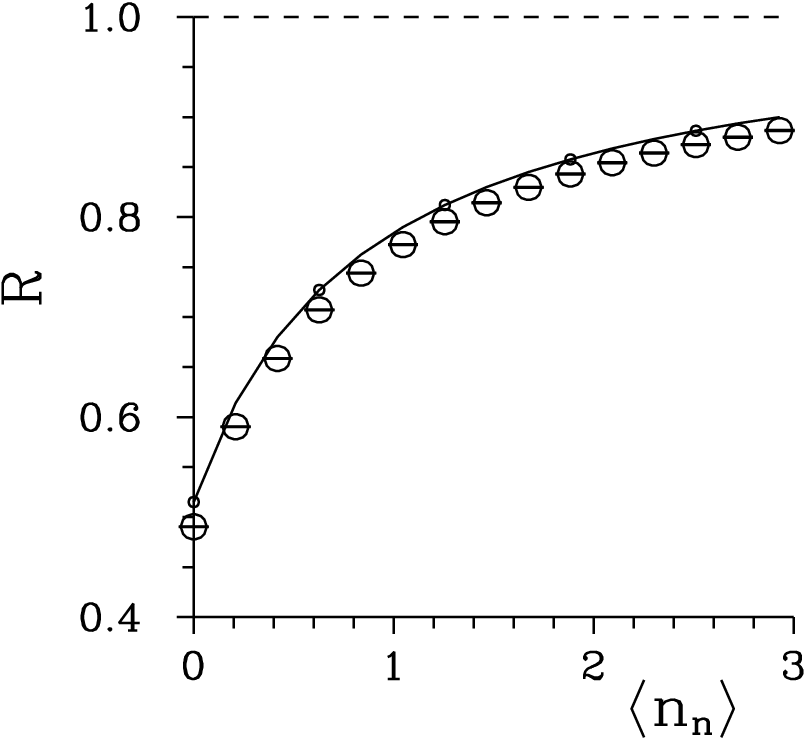}
    \hspace{2mm}
     \includegraphics[width=0.47\hsize]{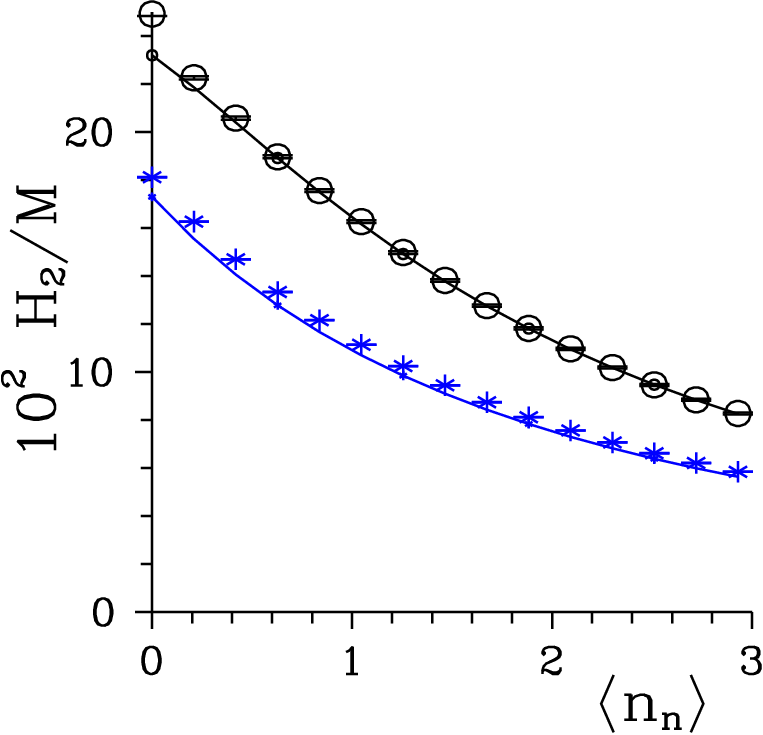}}
  \centerline{(c)\hspace{0.45\hsize} (d)}
 \caption{(a) One-beam mean photon number $ \langle n\rangle $, (b) its
  Fano factor $ F $, (c) 2-beam noise-reduction parameter $ R $, and
  2-beam Kullback-Leibler divergence per one typical spatio-temporal mode $ H_2/M $ as they depend
  on 1-beam mean noise photon number $ \langle n_{\rm n} \rangle $.
  Used symbols and curves are described in the caption to
  Fig.~\ref{fig5}. In (b) [(c)], the horizontal
  dashed line $ F =1 $ [$ R = 1 $] defines the classical-quantum border.}
\label{fig4}
\end{figure}

The applied technique provided us with 30 compound beams each
having around 0.8 mean correlated photons per beam and mean noise
photon numbers per beam increasing from 0 to 6. Their main
characteristics are quantified using 1-beam mean photon number $
\langle n\rangle $ and Fano factor $ F $, 2-beam noise-reduction
parameter $ R $, and 2-beam Kullback-Leibler divergence per mode $
H_2/M $. The 1-beam mean photon number $ \langle n\rangle $ is
plotted in Fig.~\ref{fig4}(a) for the 1-beam mean noise photon
number $ \langle n_{\rm n} \rangle $ increasing from 0 to 3.
Increasing $ \langle n_{\rm n} \rangle $, the Fano factor $ F $ of
a beam, $ F_1 = 1+ \langle (\Delta W_1)^2\rangle / \langle
W_1\rangle $, is kept roughly constant [see Fig.~\ref{fig4}(b)].
The increase of noise also relatively weakens the pairwise beams
correlations contained in the correlated units, which are the
source of entanglement of the whole state. Two-beam quantum
correlations are quantified in Fig.~\ref{fig4}(c) by the
noise-reduction parameter $ R $ defined as $ R_{12} =  1 + \langle
[ \Delta(W_1-W_2)]^2 \rangle / (\langle W_1\rangle +\langle
W_2\rangle ) $. For the analyzed states, the noise-reduction
parameter $ R $ increases from 0.5 to 0.9 because of the noise.
The increase of mean noise photon number $ \langle n_{\rm
n}\rangle $ also results in lowering inseparability of the 2-beam
reduced states [see Fig.~\ref{fig4}(d)]. Inseparability of the
state is quantified by the two-beam Kullback-Leibler divergence $
H_2 \equiv S_{{\rm R},1} + S_{{\rm R},2} - S_{{\rm R},12} $
defined as the difference between the 1-beam and 2-beam Renyi
entropies [$ S_{\rm R} = - \ln(\mu) $]. We note that, according to
Fig.~\ref{fig4}(d), inseparability of the states as determined
by the intensity moments up to sixth order is considerably larger
than the idealized one derived only from second-order intensity
moments (see below).

To discuss quantum correlations, we determine up-to sixth-order
intensity moments reduced per mode in
each beam (for the reduction, see Sec.~III in SM~\cite{SM}), derive the fields' 1-, 2-, and
3-beam purities $ \mu_1 $, $ \mu_2 $, and $ \mu_3 $, and implement
the above theoretical formulas. However, before analyzing the
experimental data, we have to address the problem of experimental
precision required to reliably reveal the needed purities. This
leads us to three approaches to the data analysis that differ by the
number of required measurement realizations.

The usual number of measurement realizations gives the intensity
moments up to fourth order with sufficient precision. This gives
the purities $ \mu_1 $ and $ \mu_2 $ and constraints on
possible values of the seralian $ \Delta_2 $. The purities $ \mu_1
$ and $ \mu_2 $ identify in the single-mode theoretical model a
group of compatible STBGSs whose analysis sets the lower (green
curves and symbols) and upper (red curves and symbols) bounds for
the seralian $ \Delta_2 $. Alternatively, we can estimate the
value of seralian $ \Delta_2 $ directly from the experimental
intensity moments (see Eq.~\eqref{eq:purity3} and Appendix~A).
Nevertheless, the obtained high-quality data ($ 6.95\times 10^8 $
measurements on weak twin beams provided $ \approx 6.95/6\times
10^8 $ realizations of the analyzed states) allow us to reliably
determine the intensity moments up to sixth order. They uniquely
identify the state and so give all fields' characteristics (black
curves and symbols). On the other hand, having technical
applications in mind and assuming the specific form of STBGSs
imposed by the used weak twin beams, the second-order intensity moments
suffice in the state identification (dark blue curves and symbols,
see \cite{Arkhipov2015}). Our experimental data suggest that about
$ 10^8 $ measurements are needed to completely identify the state
(for more details, see Appendix~B). About $ 10^6 $ measurements
allow us to safely characterize the state using the lower and
upper bounds through fourth-order intensity moments. Finally, only about $ 10^4 $ state realizations
suffice when we rely only on second-order intensity moments. We
note that the analyzed states can directly be emitted in the
experiment once the APDs are replaced by photon-number-resolving
detectors (see, e.g., in \cite{Michalek2020}).

We note here that the structure of the experimental fields is
richer than that of the single-mode theoretical model and so the
above three approaches to the experimental data analysis do not
necessarily give compatible predictions for the fields'
properties. The more elaborated approach is used, i.e. the higher
the intensity moments are involved, the more reliable and more
detailed the predictions are. However, the experimental errors
must be kept reasonably low.

Considering the intensity moments $ \langle w_1^{k_1} w_2^{k_2}
w_3^{k_3}\rangle $ reduced per one typical spatio-temporal mode in
each beam, the formula for 1-beam purity $ \mu_1 $ is simple,
\begin{equation} 
 \mu_1 = \frac{1}{ \sqrt{ 1+ 4\langle w_1\rangle
   + 12\langle w_1\rangle^2 - 4\langle w_1^2\rangle } }.
\label{A6}
\end{equation}
Contrary to this, the formulas for 2- and 3-beam purities $ \mu_2 $ and $ \mu_3 $  are
rather complex and they can be found in \cite{BarasinskiPRL2023}
and \cite{Sudak2025}, respectively.

\begin{figure}  
  \centerline{\includegraphics[width=0.47\hsize]{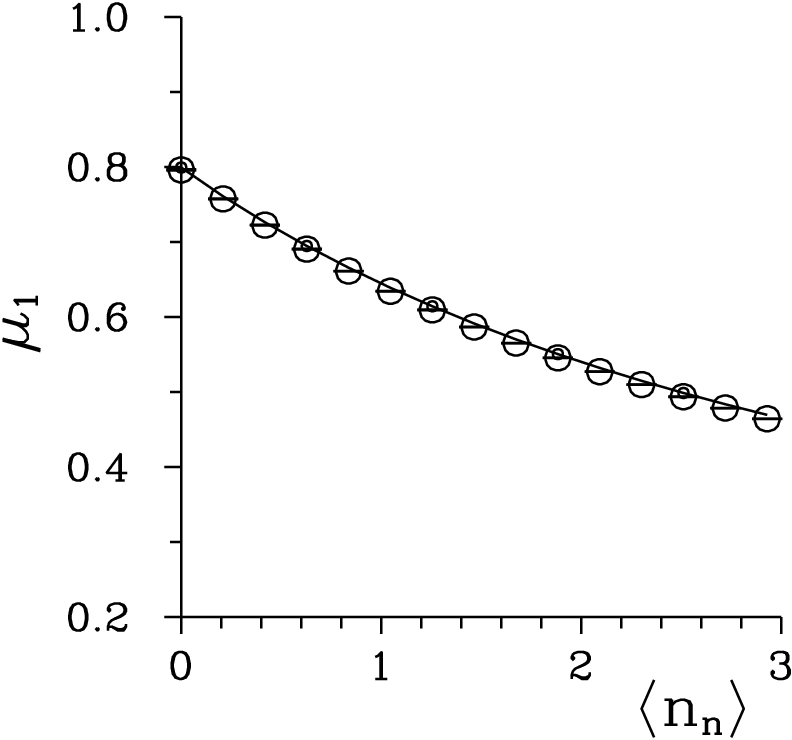}
    \hspace{2mm}
     \includegraphics[width=0.47\hsize]{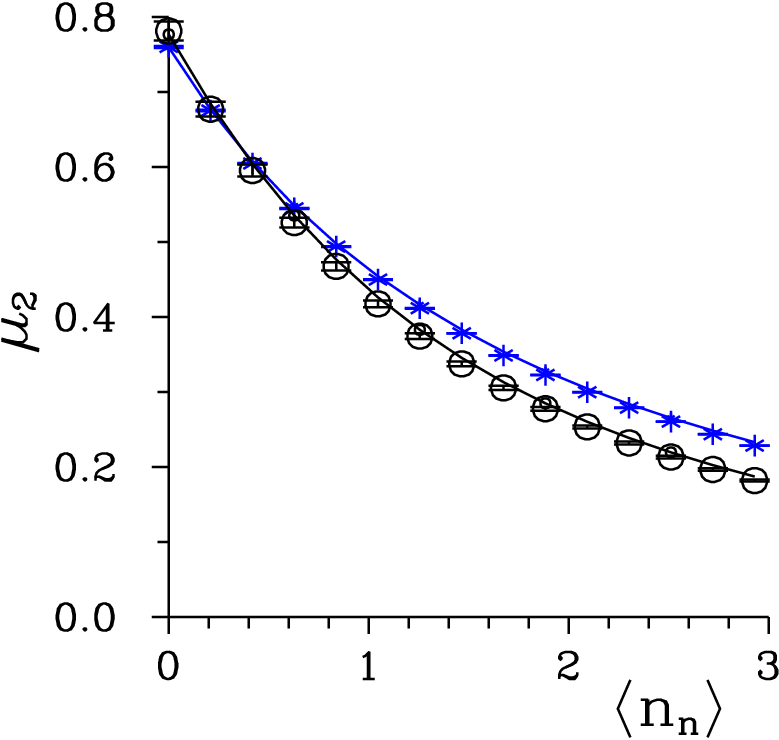}}
  \centerline{(a)\hspace{0.45\hsize} (b)}

  \centerline{\includegraphics[width=0.47\hsize]{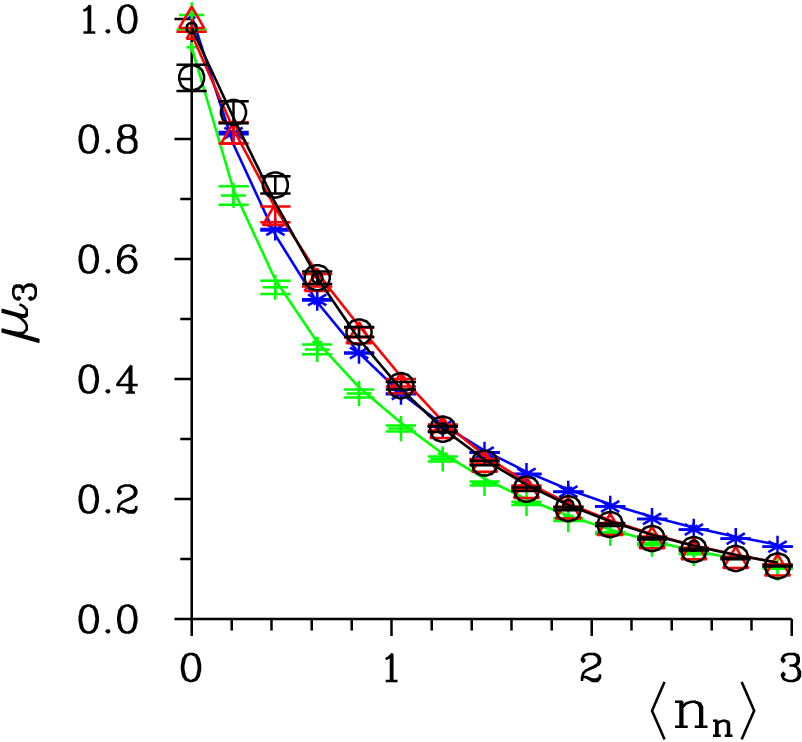}
    \hspace{2mm}
     \includegraphics[width=0.47\hsize]{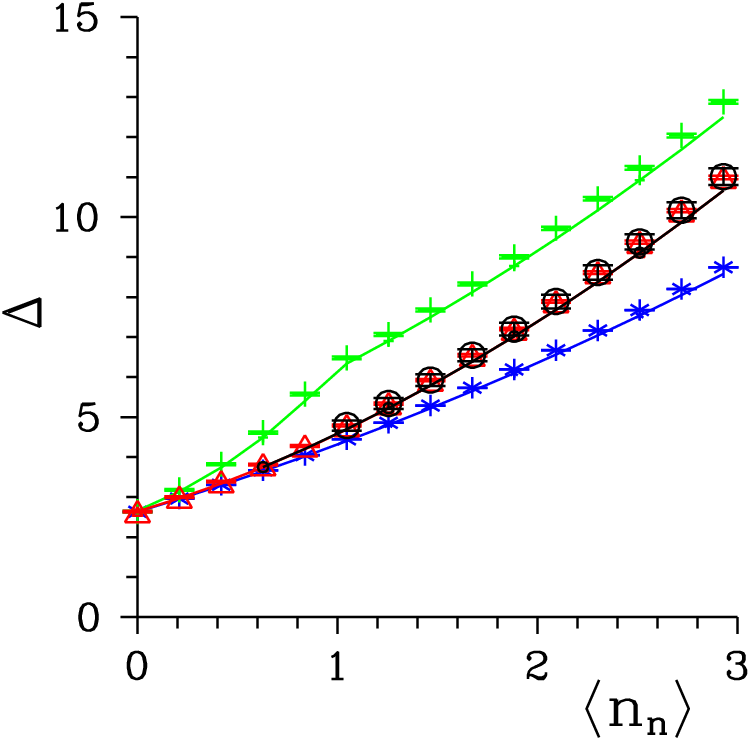}}
  \centerline{(c)\hspace{0.45\hsize} (d)}

 \caption{(a) One-, (b) two-, and
  (c) 3-beam purities $ \mu_1 $, $ \mu_2 $, and $ \mu_3 $ and (d) seralian $ \Delta_2 $ as they depend on 1-beam mean noise photon number $ \langle n_{\rm n} \rangle $.
  Isolated symbols with error bars give experimental data, curves
  originate in the general Gaussian model.
  Black curves and symbols $ \circ $ denote {\emph exact} values obtained
  from the intensity moments up to sixth order.
  The use of moments only up to second order gives blue
  curves and symbols $ \ast $. Fourth-order moments provide the
  lower (green curves and symbols +) and upper (red curves and symbols $ \triangle $)
  estimates for the determined quantities.}
\label{fig5}
\end{figure}

In the analyzed experiment described in detail in
Ref.~\cite{PerinaJr.2021}, the detectors were characterized by
detection efficiencies $ \eta_{\rm s} = 0.274 $, $ \eta_{\rm i} =
0.324 $ and dark-count rates $ d_{\rm s} = 2.8 \times 10^{-3} $,
$ d_{\rm i} = 3.8 \times 10^{-3} $. The application of the above
mentioned formulas gave us the purities $ \mu_1 $, $ \mu_2 $, and
$ \mu_3 $ as shown in Fig.~\ref{fig5}.
They naturally decrease with the increasing mean noise photon
number $ \langle n_{\rm n}\rangle $. According to the graphs in
Figs.~\ref{fig5}(a,b,c), we have $ \mu_1 > \mu_2 > \mu_3 $ for
all experimental STBGSs.  Moreover, as
the seralian $ \Delta_2 $ is an important parameter for
characterizing the state, we apply Eqs.~(\ref{eq:delta3}) and
(\ref{eq:p1}) to determine its values from those of the purities
[see Fig.~\ref{fig5}(d)].

\begin{figure}  
  \centerline{\includegraphics[width=0.47\hsize]{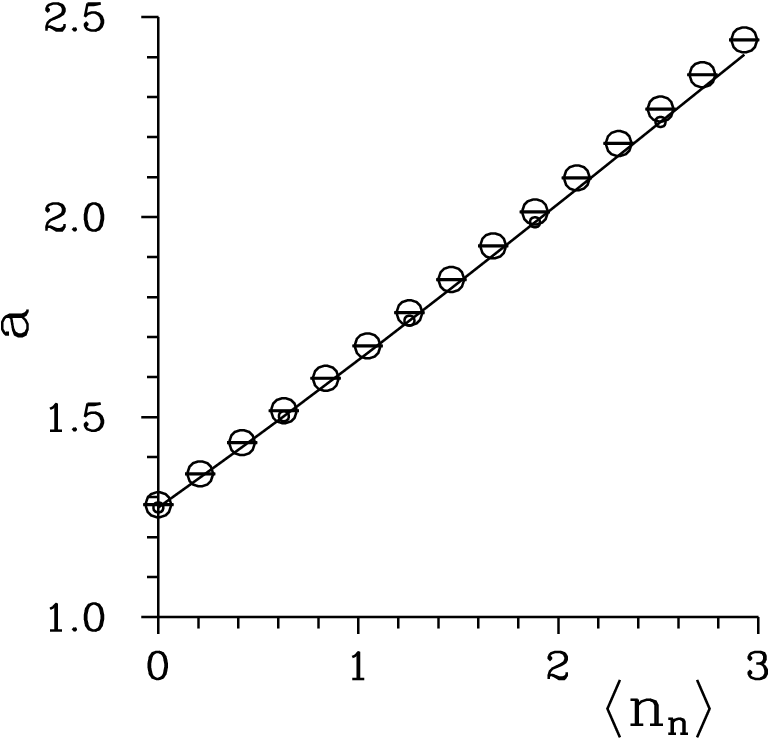}
    \hspace{2mm}
     \includegraphics[width=0.47\hsize]{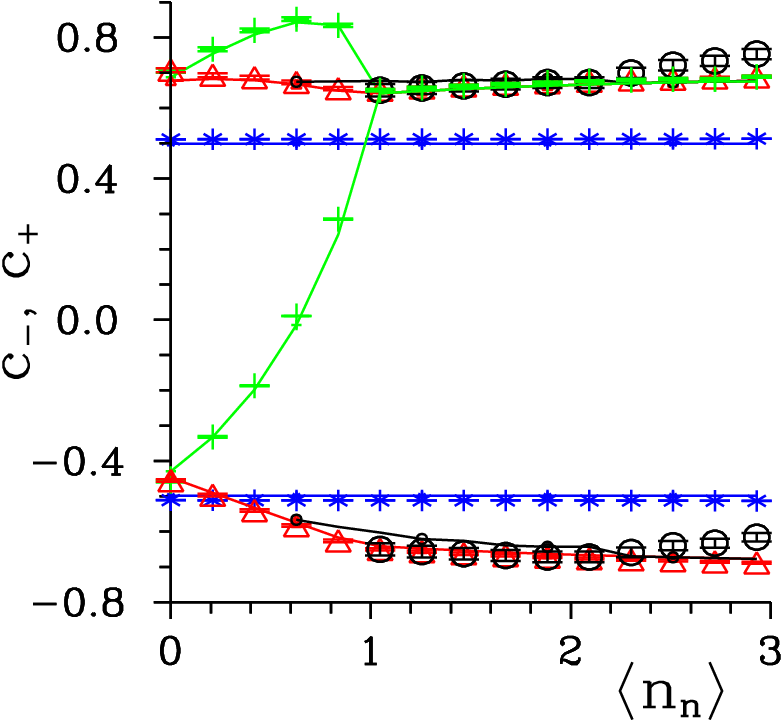}}
  \centerline{(a)\hspace{0.45\hsize} (b)}
 \caption{State coefficients (a) $ a $ and (b) $ c_- $ and $ c_+ $, $ c_- \le c_+ $,
  as they depend on 1-beam mean noise photon number $ \langle n_{\rm n} \rangle $.
  Symbols and curves are described in the caption to
  Fig.~\ref{fig5}. }
\label{fig6}
\end{figure}

The formula \eqref{eq:p1} also allows us to derive the coefficients
$ a $, $ c_+ $, and $ c_- $ from the above purities, as shown in
Fig.~\ref{fig6}.
To arrive at the values of coefficients $ c_+ $ and $ c_- $, we
need to know the 3-beam purity $ \mu_3 $, which is experimentally
demanding. Alternatively, we may rely only on the experimental
values of purities $ \mu_1 $ and $ \mu_2 $ and set the lower and
upper bounds for the coefficients $ c_+ $ and $ c_- $.

\section{Observed quantum correlations: Genuine tripartite entanglement}

\begin{figure}  
  \centerline{\includegraphics[width=0.47\hsize]{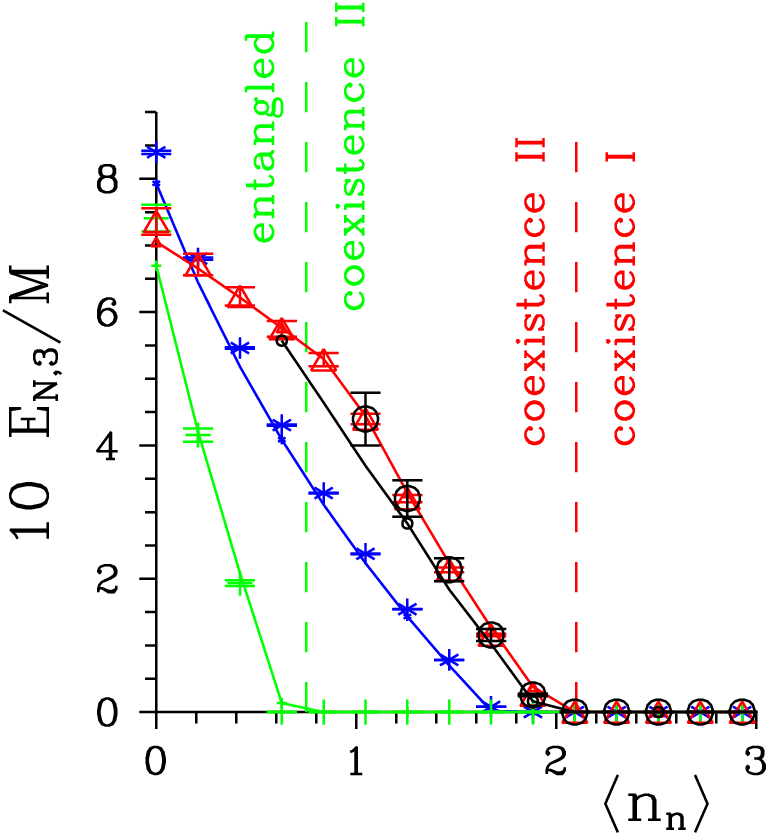}
    \hspace{2mm}
     \includegraphics[width=0.47\hsize]{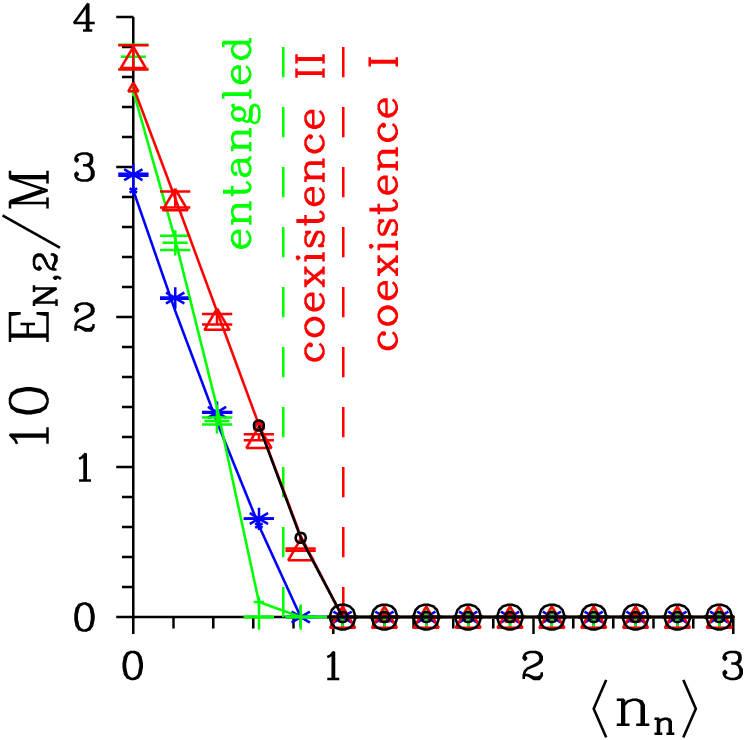}}
  \centerline{(a)\hspace{0.45\hsize} (b)}
  \centerline{\includegraphics[width=0.47\hsize]{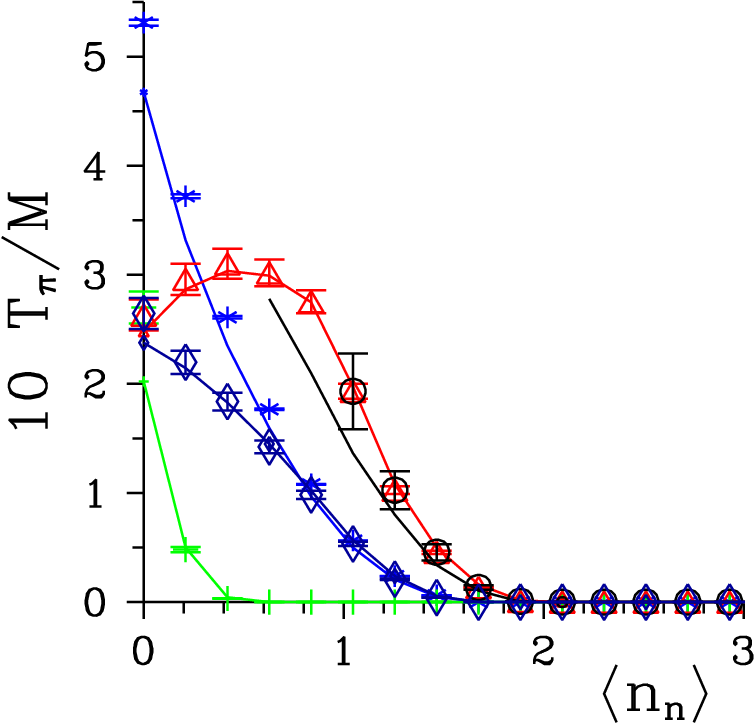}
    \hspace{2mm}
     \includegraphics[width=0.47\hsize]{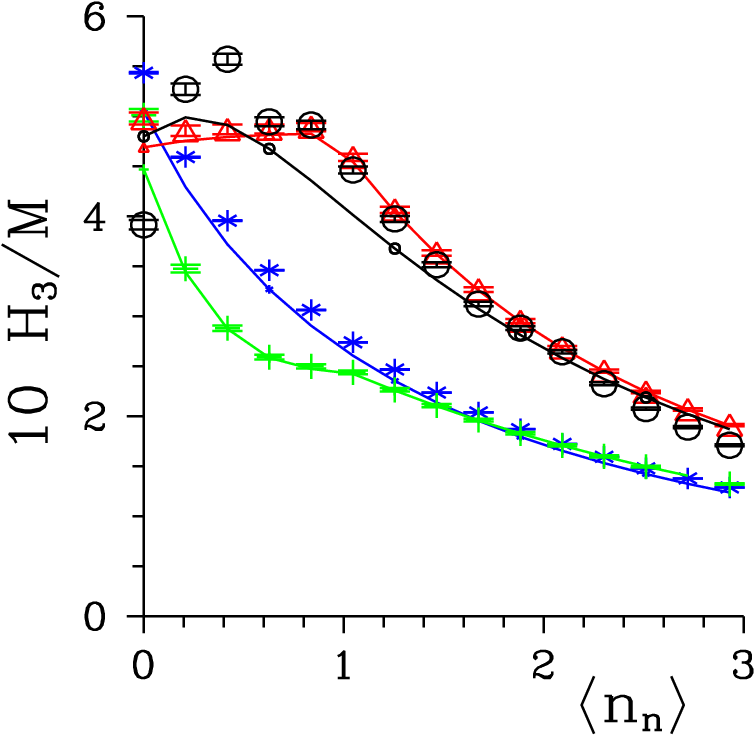}}
  \centerline{(c)\hspace{0.45\hsize} (d)}
 \caption{(a) Tripartite negativity per mode $ E_{N,3}/M $, (b) bipartite negativity per mode $ E_{N,2}/M $,
   (c) cotangle per mode $ T_\pi/M $, and (d) 3-beam Kullback-Leibler divergence per mode $ H_3/M $ as they
  depend on 1-beam mean noise photon number $ \langle n_{\rm n} \rangle $.
  Symbols and curves are described in the caption to
  Fig.~\ref{fig5}.
  In (a) and (b), the point where the
  green (red) curve  nullifies gives the border between entangled
  (coexistence II) and coexistence II (coexistence I) regions
  defined in Table \ref{tab_1}. In (c), navy blue curve and symbols
  $ \diamond $ refer to noisy GHZ/W states identified by
  $ \mu_1 $ and $ \mu_2 $, as discussed in Sec.~VII below.}
\label{fig7}
\end{figure}

The 3-beam correlations in the analyzed states are established by
2-beam correlations (using photon pairs) covering all three beams.
This results in the fields' 3-beam entanglement (tripartite
entanglement $A-BC$), quantified by the logarithmic negativity $
E_{N,3} $, that naturally decreases with the increasing mean noise
photon number $ \langle n_{\rm n} \rangle $ [see
Fig.~\ref{fig7}(a)].
The simplest approach for data processing as well as the most
elaborated one suggest that the tripartite entanglement disappears
for $ \langle n_{\rm n}\rangle \approx 2 $, i.e. when the noise is
approximately 2.5-times larger than the correlated signal.
Tripartite entanglement includes contributions from both bipartite
entanglement (between beams $A-B$ and $A-C$) and genuine
tripartite entanglement. By quantifying bipartite entanglement
using the logarithmic negativity $ E_{N,2}$, the genuine
tripartite entanglement can be inferred by calculating the
residual cotangle \cite{Adesso_NJP2006} introduced in
Eq.~(\ref{cotangle}). Following the curves in Fig.~\ref{fig7}(b),
the bipartite negativity $ E_{N,2} $ is lost already for $ \langle
n_{\rm n} \rangle $ around 1. As shown in Figs.~\ref{fig7}(a,b,c),
the fields with $ \langle n_{\rm n}\rangle $ below 2 exhibit the
genuine tripartite entanglement. In particular, for $ \langle
n_{\rm n}\rangle $ between 1 and 2, this entanglement persists
even in the absence of the bipartite entanglement. Interestingly,
inside this region, the values of cotangle $T_\pi$ are
significantly lower compared to those observed for $ \langle
n_{\rm n}\rangle <1 $, which highlights the promiscuous nature of
CV entanglement sharing \cite{Adesso_NJP2006}. We note that the
properties of the analyzed states are close to those of the noisy
GHZ/W states, as evidenced by the corresponding values of cotangle
$T_\pi$ in Fig.~\ref{fig7}(c) (for details, see Sec.~VII below).
The approach based on up to fourth-order intensity moments
confirms this by determining the upper and lower bounds of these
quantities, though it guarantees the genuine tripartite
entanglement only for the fields with  $ \langle n_{\rm n} \rangle
$ smaller than 0.4. We note that, according to
Figs.~\ref{fig7}(a,b), tripartite entanglement is more resistant
against the noise than the bipartite entanglement. Inseparability
of 3-beam fields to their one- and two-beam counterparts,
quantified by 3-beam Kullback-Leibler divergence $ H_3 \equiv
S_{{\rm R},A} + S_{{\rm R},BC} - S_{{\rm R},ABC} $ defined in
terms of the R\'{e}nyi-2 entropies $ S_{\rm R} $ and plotted in
Fig.~\ref{fig7}(d), naturally decreases with the increasing $
\langle n_{\rm n}\rangle $.

\begin{figure}  
  \centerline{\includegraphics[width=0.47\hsize]{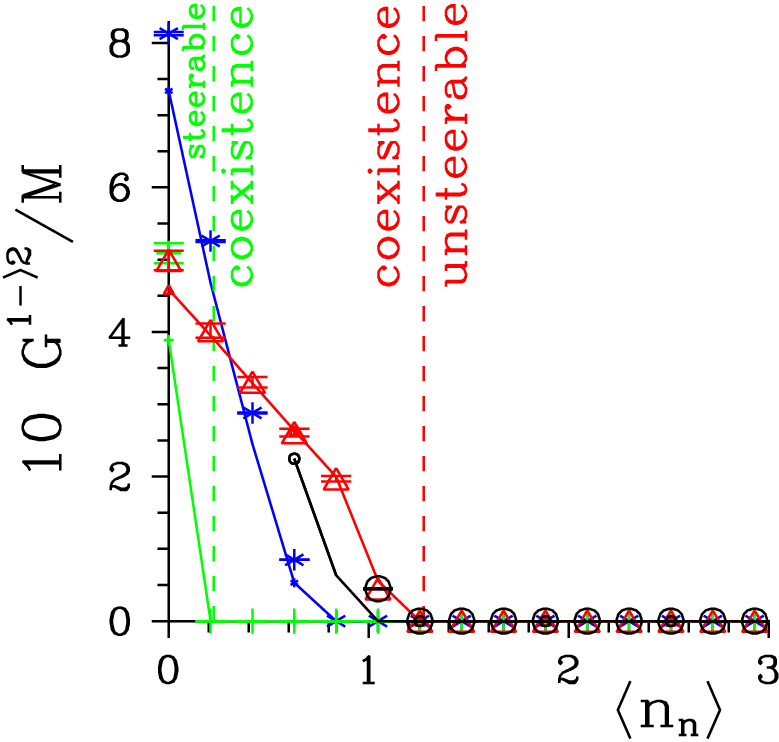}
    \hspace{2mm}
     \includegraphics[width=0.47\hsize]{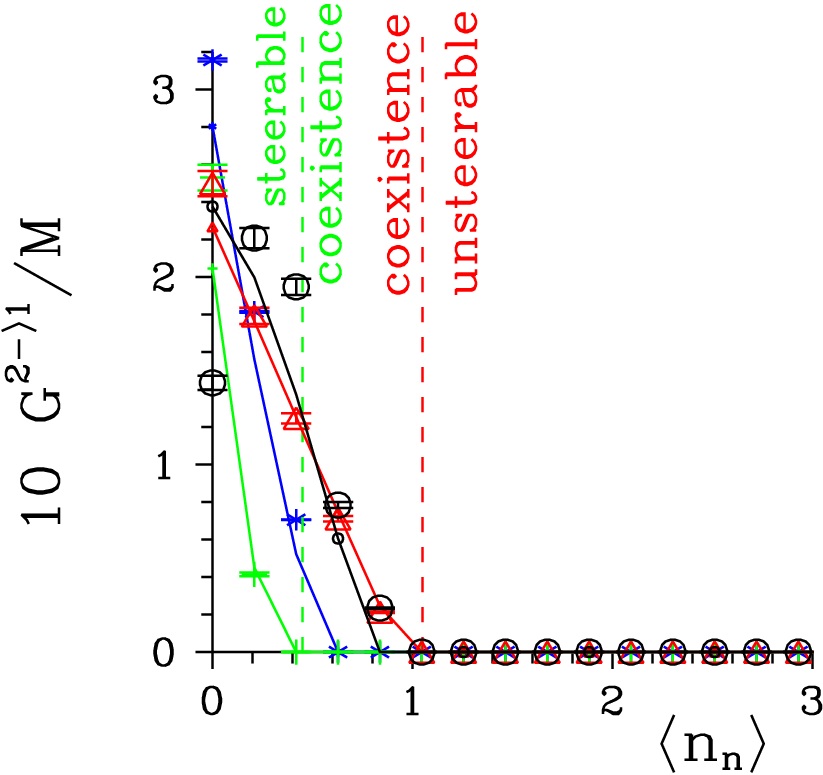}}
  \centerline{(a)\hspace{0.45\hsize} (b)}
 \caption{Three-beam steering parameters per mode (a) $ G^{1\rightarrow 2}/M $ and (b)
  $ G^{2\rightarrow 1}/M $ as they
  depend on 1-beam mean noise photon number $ \langle n_{\rm n} \rangle $.
  Used symbols and curves are described in the caption to
  Fig.~\ref{fig5}. The point where the green (red) curve nullifies identifies the border
  between steerable (coexistence) and coexistence (unsteerable) regions defined in Table \ref{tab_1}.}
\label{fig8}
\end{figure}

Steering represents a stronger form of quantum correlations. For
our STBGSs, 2-beam steering does not
occur. On the other hand, both one beam can steer the remaining
two beams (steering parameter $ G^{1\rightarrow 2} $) and vice
versa ($ G^{2\rightarrow 1} $), as evidenced and quantified in
Figs.~\ref{fig8}(a,b). Steering by one beam is stronger than its
two-beam counterpart, and they are both observed for the states
with low amount of the noise $ \langle n_{\rm n}\rangle < 1 $.

\section{Comparison with theoretical noisy GHZ/W states}

\begin{figure}  
  \centerline{\includegraphics[width=0.47\hsize]{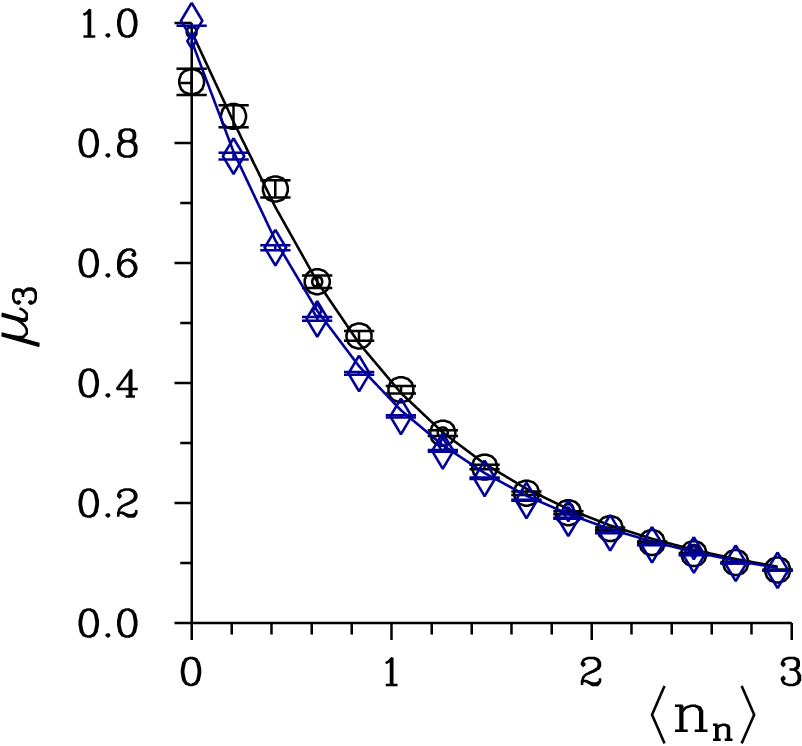}
    \hspace{2mm}
     \includegraphics[width=0.47\hsize]{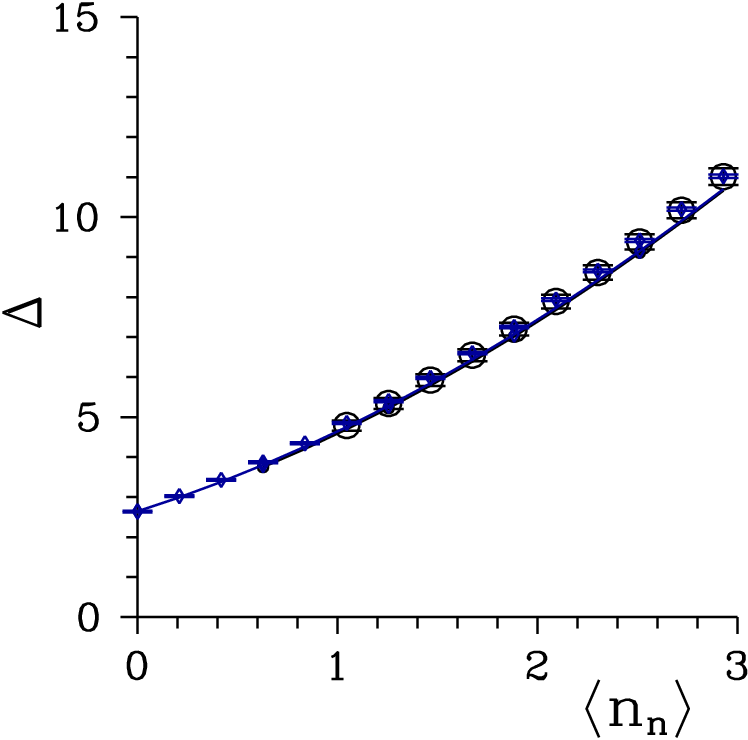}}
  \centerline{(a)\hspace{0.45\hsize} (b)}

  \centerline{\includegraphics[width=0.47\hsize]{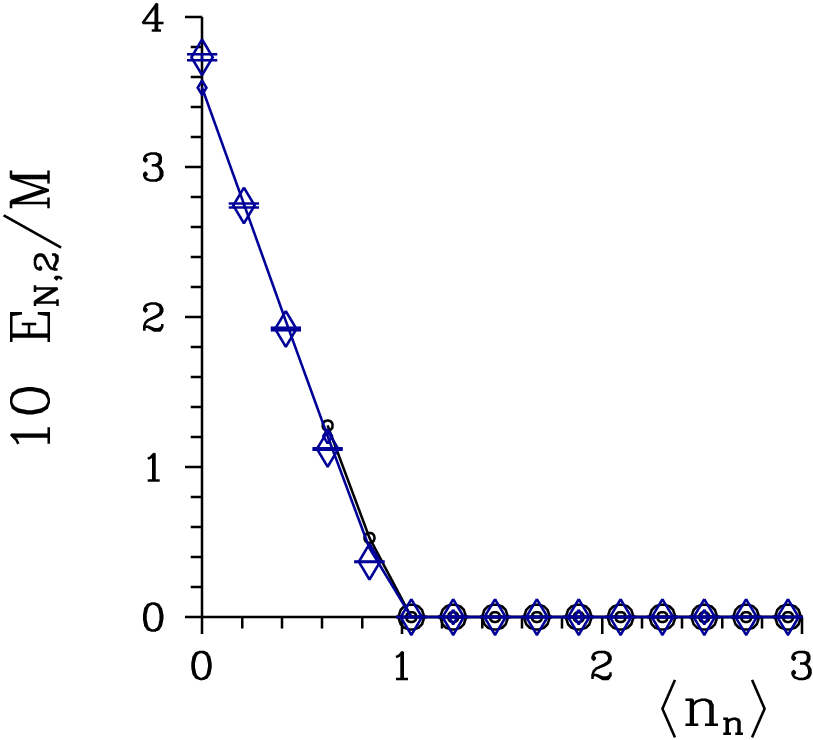}
    \hspace{2mm}
     \includegraphics[width=0.47\hsize]{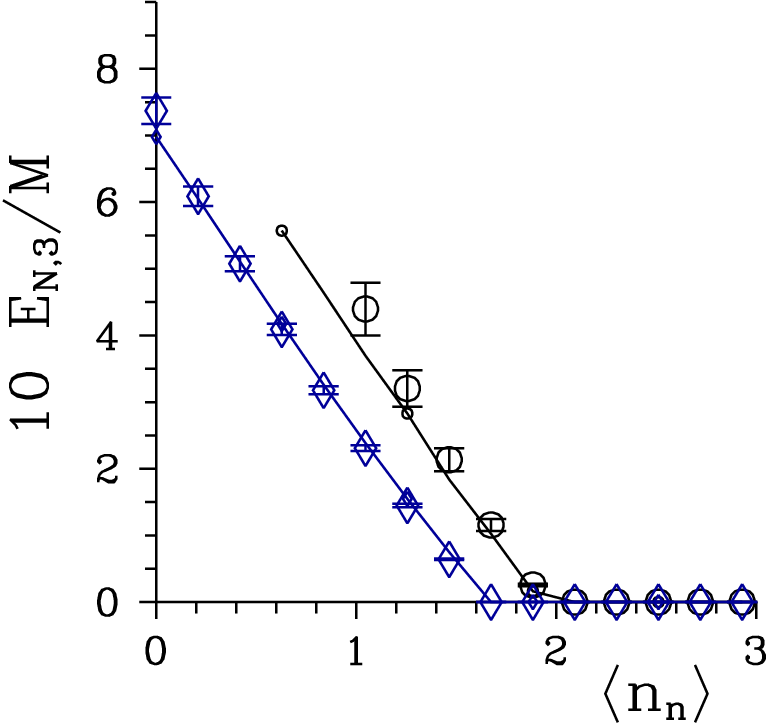}}
  \centerline{(c)\hspace{0.45\hsize} (d)}

  \centerline{\includegraphics[width=0.47\hsize]{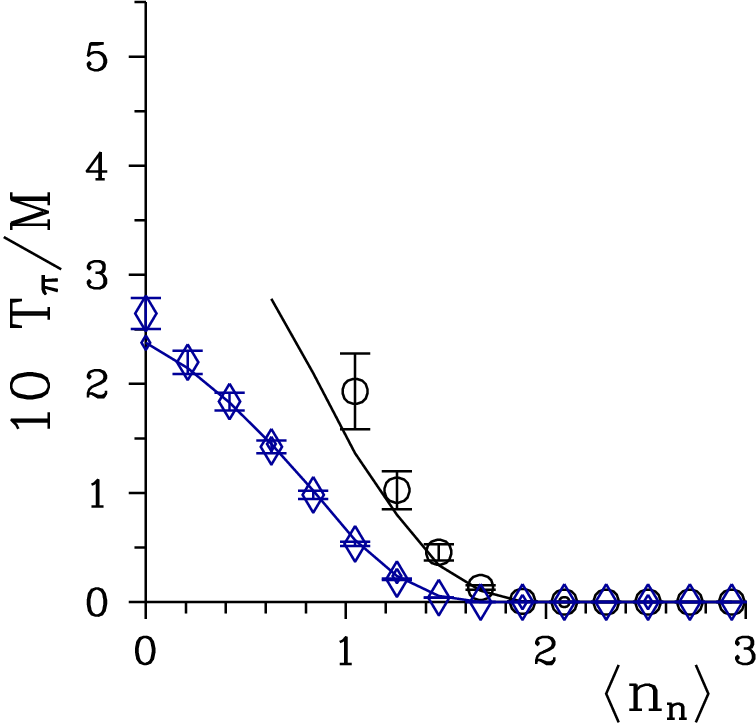}
    \hspace{2mm}
     \includegraphics[width=0.47\hsize]{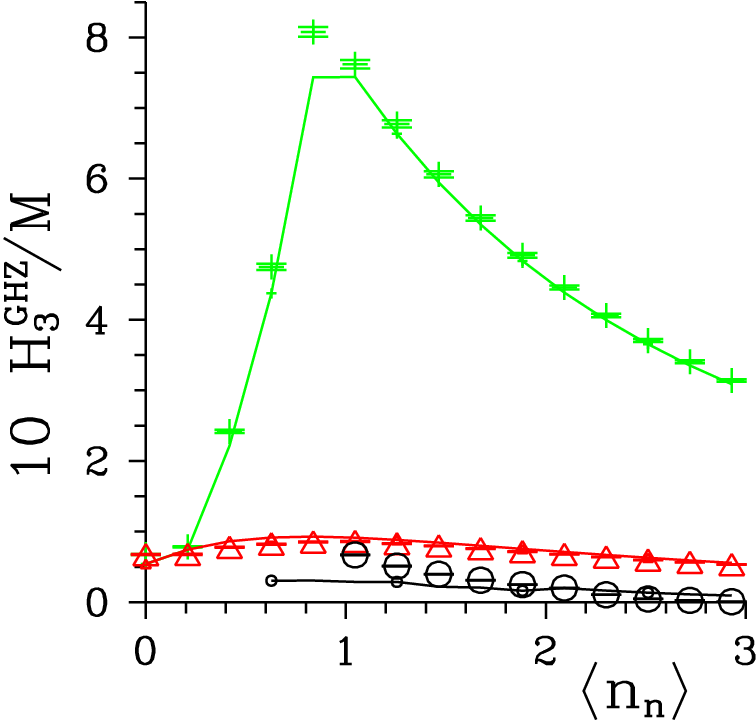}}
  \centerline{(e)\hspace{0.45\hsize} (f)}

 \caption{(a) 3-beam purity $ \mu_3 $, (b) seralian $ \Delta_2 $, (c) 2-beam negativity per mode $ E_{N,2}/M $,
  (d) 3-beam negativity per mode $ E_{N,3}/M $, (e) cotangle per mode $ T_\pi/M $, and
  (f) 3-beam Kullback-Leibler divergence per mode $ H_3^{\rm GHZ}/M $ from
  the corresponding GHZ/W states  as they depend on 1-beam mean noise photon
  number $ \langle n_{\rm n} \rangle $.
  For GHZ/W states: navy blue curves originate in the general Gaussian
  model, isolated navy blue symbols $ \diamond $ with error bars give experimental
  data. The remaining symbols and curves are described in the caption to
  Fig.~\ref{fig5}; $ M=6.7 $.}
\label{fig9}
\end{figure}

In the last section, we provide detailed comparison of the
properties of the generated states with those of the noisy three-mode GHZ/W states. These states belong to the family of
fully symmetric mixed Gaussian states, also known as three-mode
squeezed thermal states \cite{CHEN2005121}. The properties of
these states were extensively studied in the literature
\cite{Adesso2007}. The noisy GHZ/W states are characterized by two
independent parameters, which define a fully symmetric covariance
matrix $\sigma_3$ that was introduced in Eq.~(\ref{eq:CM3}). Adopting
the parametrization from Ref.~\cite{Adesso2007}, the following relations
are derived:
\begin{eqnarray}  
\mu_3 &=& \left(\frac{\mu_2}{\mu_1}\right)^3,\nonumber\\
\Delta_2 &=& \frac{1}{\mu_1^2} + \frac{\mu_1^2}{\mu_2^2}.
\label{G1}
\end{eqnarray}
The values of purity $ \mu_3 $ and seralian $ \Delta_2 $ of the
GHZ/W states are compared with those of the generated states in
Figs.~\ref{fig9}(a) and (b), respectively. Whereas the purities
$ \mu_3 $ of GHZ/W states are close to the upper bound of $ \mu_3
$ of the generated states from the bottom, the seralians $
\Delta_2 $ are close to the lower bound of $ \Delta_2 $ of the
generated states from the upper. This means that the GHZ/W states
belong the the states with the greatest entanglement among the
states sharing the purities $ \mu_1 $ and $ \mu_2 $. The
corresponding values of 2-beam negativity per mode $ E_{N,2}/M $,
3-beam negativity per mode $ E_{N,3}/M $, and cotangle per mode $
T_\pi/M $ as they vary with the increasing mean noise photon
number $ \langle n_{\rm n} \rangle $ are plotted by navy blue symbols $
\diamond$ and curves in turn in Figs.~\ref{fig9}(c), (d), and
(e).

The vicinity of the GHZ/W states to the generated states is judged using the Kullback-Leibler divergence $ H_3^{\rm GHZ} $ that is
determined for the generated states and the GHZ/W states along the formula:
\begin{eqnarray}    
H_3^{\rm GHZ} &=& \frac{ -3 \sqrt{h_1}+ h_2}{8 \mu_1^4 \mu_2} +
  \ln \left( \frac{ \sqrt{2} \mu_1^4}{ \mu_2^2\sqrt{h_3} } \right), \\
 h_1 &=& [-4 \mu_1^4 + (-4 + \Delta_2 \mu_1^2)^2 \mu_2^2] [\mu_1^8
    - 10 \mu_1^4 \mu_2^2 + 9\mu_2^4] , \nonumber \\
 h_2 &=& 3 (-\mu_1^4 + \mu_2^2) \left(8 \mu_2 + \mu_1^2 \sqrt{-4 + \Delta_2^2
    \mu_2^2}\right), \nonumber \\
 h_3 &=& 6 + \Delta_2^2 (-3 + \Delta_2 \mu_1^2) \mu_2^2
        - 2 \mu_1^2 \sqrt{-4 + \Delta_2^2 \mu_2^2} /\mu_2
        \nonumber \\
  & & \hspace{5mm} + \Delta_2 (3 - \Delta_2 \mu_1^2) \mu_2 \sqrt{-4 + \Delta_2^2 \mu_2^2}
\label{G2}
\end{eqnarray}
As evidenced by the values of the Kullback-Leibler divergence per
mode $ H_3^{\rm GHZ}/M $ plotted in Fig.~\ref{fig9}(f) the
generated states and the corresponding GHZ/W states are rather
close to each other: The greater the mean noise photon number $
\langle n_{\rm n} \rangle $ is, the closer the states are.

Similarity of the two kinds of states suggests to apply the
entanglement classification developed for the GHZ/W states in
\cite{CHEN2005121, Adesso2007} to the generated states. The
corresponding classification is written as
\begin{eqnarray}  
   & \sqrt{\frac{\mu_1^3}{2 - \mu_1}} < \mu_2 \rightarrow \textbf{class 1,} & \nonumber\\
   & \sqrt{\frac{5 \mu_1^4 + 3 \sqrt{\mu_1^6 (8 + \mu_1^2)}}{18 - 4 \mu_1^2}} < \mu_2 \leq \sqrt{\frac{\mu_1^3}{2 - \mu_1}} \rightarrow \textbf{class 4,} & \nonumber\\
   & \mu_1^2 < \mu_2 \leq \sqrt{\frac{5 \mu_1^4 + 3 \sqrt{\mu_1^6 (8 + \mu_1^2)}}{18 - 4 \mu_1^2}}
   \rightarrow  \textbf{class 5.} &
\label{G3}
\end{eqnarray}
According to this classification the corresponding noisy GHZ/W
states fall into class 1 for $\langle n_{\rm n} \rangle < 1.67$
and class 5 for $\langle n_{\rm n} \rangle > 1.88$. Class 4 was
not observed, probably owing to the lower number of
measurements in the intermediate range of $\langle n_{\rm n} \rangle $.

Furthermore, the coexistence of bi- and tri-partite CV
entanglement occurs when the conditions
$\mu_1\left(\frac{1+\mu_1^2}{3-\mu_1^2}\right)^{1/2}<\mu_2$ and
$\mu_1 < \sqrt{3} \mu_2$ are fulfilled, as discussed in
\cite{AdessoNPJ2007}. According to our experimental data, this regime is observed for $\langle n_{\rm n} \rangle < 1.04$.

\section{Conclusion}

Three-beam symmetric Gaussian states were uniquely identified
through their universal quantum invariants -- 1-, 2-, and 3-beam
purities. These invariants were experimentally determined using
second-, fourth-, and sixth-order intensity moments obtained
through photon-number-resolving detection. The states exhibiting
genuine tripartite entanglement and the state featuring the
coexistence of two- and three-beam entanglement were observed this
way. While the generated states closely resemble the noisy GHZ/W
states, our method reveals subtle differences in their properties
that demonstrate its sensitivity and reliability. Due to its
conceptual simplicity, experimental feasibility, and strong
diagnostic power, the presented method has significant potential
for broad applications. It can be readily extended to the study of
more complex families of Gaussian states. Foreseen applications
span a variety of areas involving multipartite entanglement. This includes characterization of entanglement sources and analysis of
quantum states after propagation through various media and
communication channels. These insights are particularly relevant
for complex multimode optical fields, such as those encountered in
quantum networks, quantum communication protocols, and quantum metrology.

\acknowledgments J.P. and A.\v{C}. acknowledge support by the
project ITI CZ.02.01.01/00/23\_021/0008790  of the Ministry of
Education, Youth, and Sports of the Czech Republic and EU.


\appendix

\section{Estimates and bounds using fourth-order intensity moments}

Fourth-order intensity moments allow for the estimation of the
values of seralian $ \Delta_2 $ in two different ways. In the
first one applied in the main text, the 1- and 2-beam purities $
\mu_1 $ and $ \mu_2 $ are determined from these moments. Then,
using the single-mode theoretical model, all allowed values of the
seralian $ \Delta_2 $ are identified and the lower and upper
bounds of $ \Delta_2 $ are revealed. Then, the upper and lower
bounds of 2-beam negativity $ E_{N,2} $, 3-beam negativity $
E_{N,3} $, and cotangle $ T_\pi $ are derived.

\begin{figure*}  
  \centerline{\includegraphics[width=0.23\hsize]{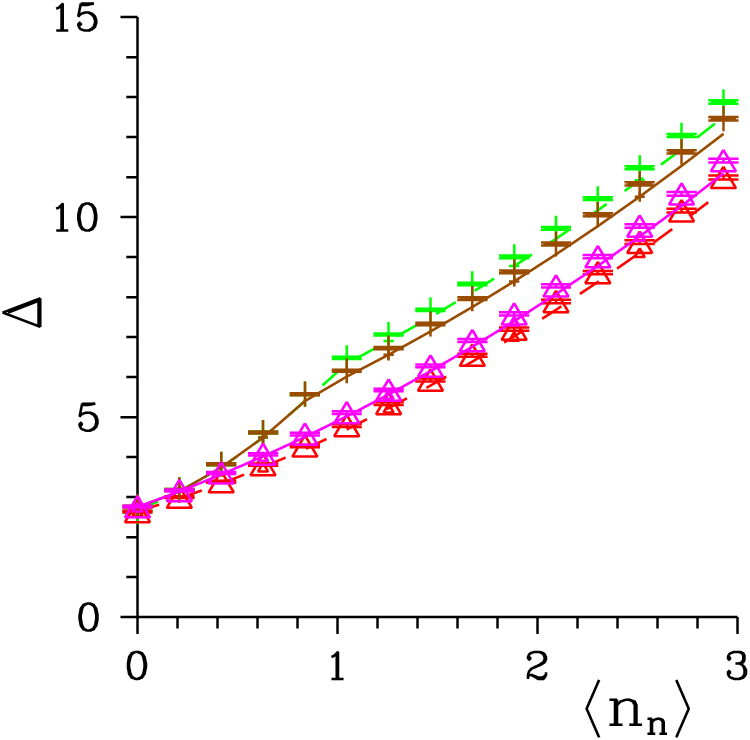}
    \hspace{2mm}
     \includegraphics[width=0.23\hsize]{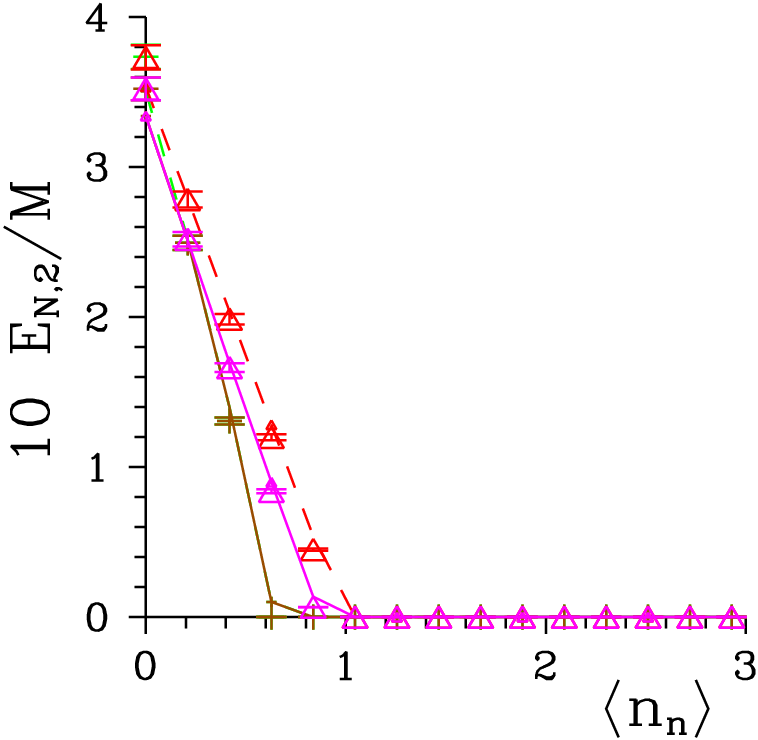}
    \hspace{2mm}
  \includegraphics[width=0.23\hsize]{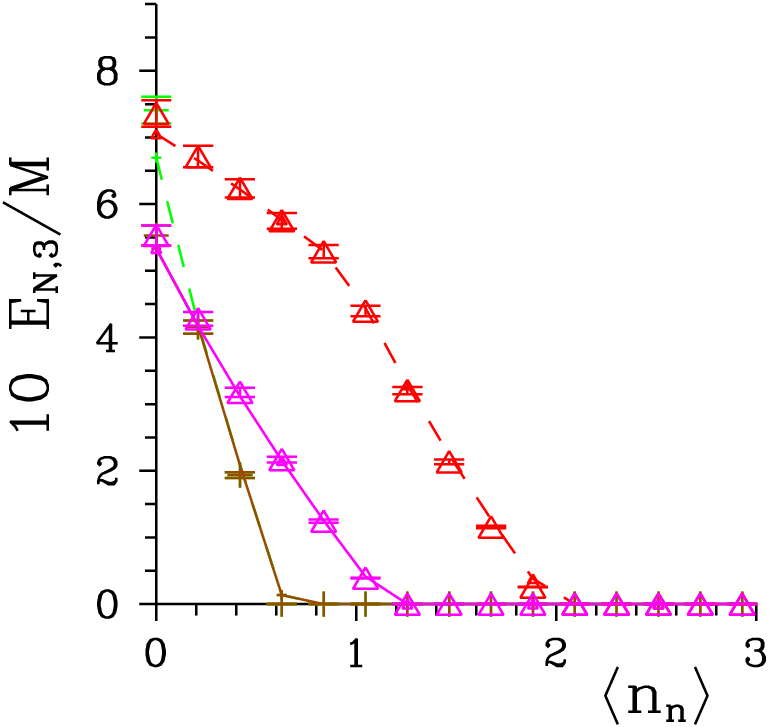}
    \hspace{2mm}
     \includegraphics[width=0.23\hsize]{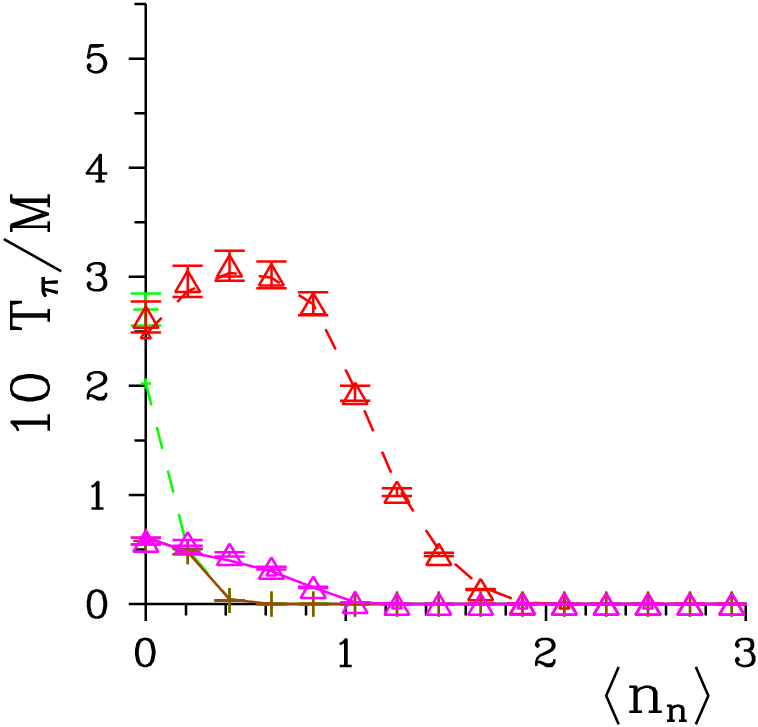}}
  \centerline{(a)\hspace{0.22\hsize} (b) \hspace{0.22\hsize}
   (c)\hspace{0.22\hsize} (d)}

  \centerline{\includegraphics[width=0.23\hsize]{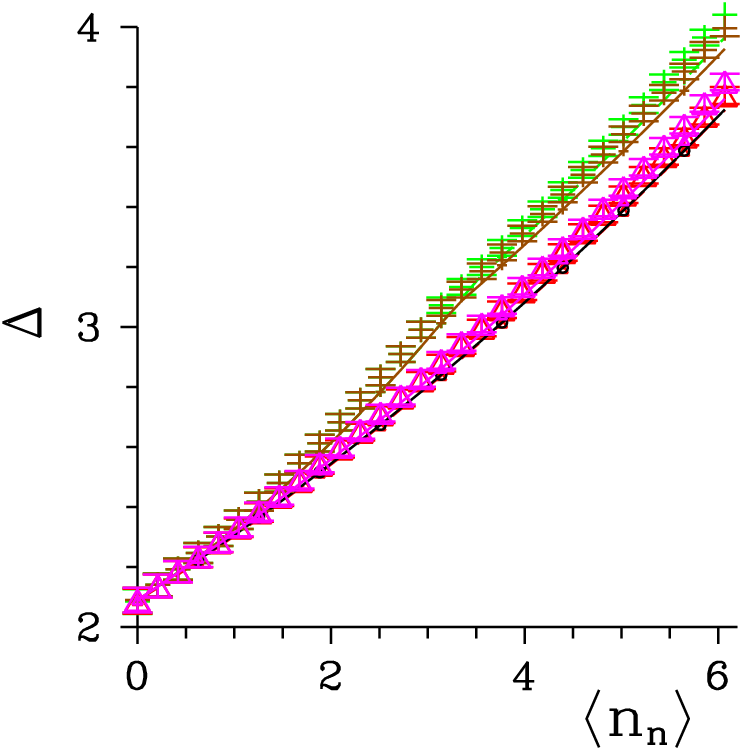}
    \hspace{2mm}
     \includegraphics[width=0.23\hsize]{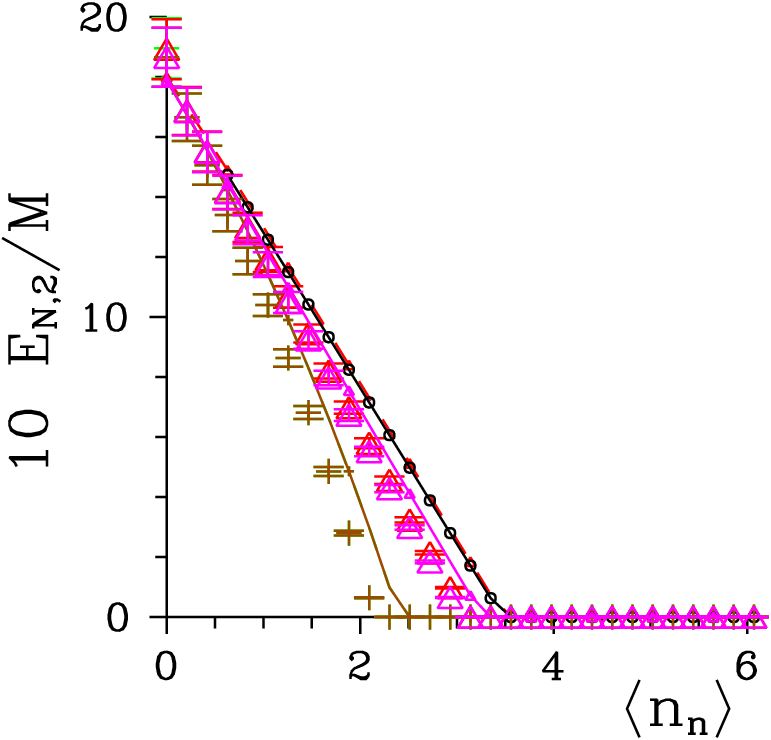}
    \hspace{2mm}
     \includegraphics[width=0.23\hsize]{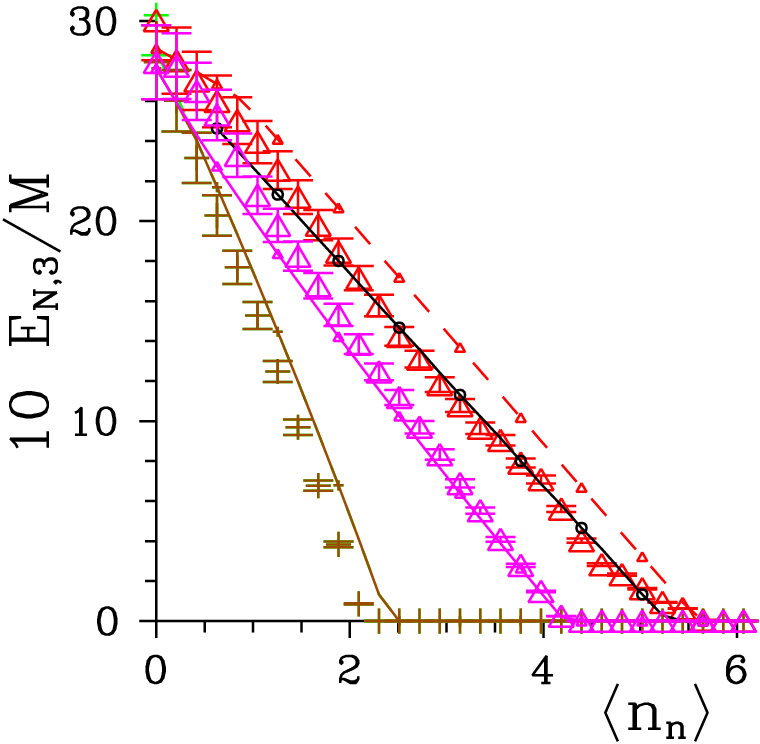}
    \hspace{2mm}
     \includegraphics[width=0.23\hsize]{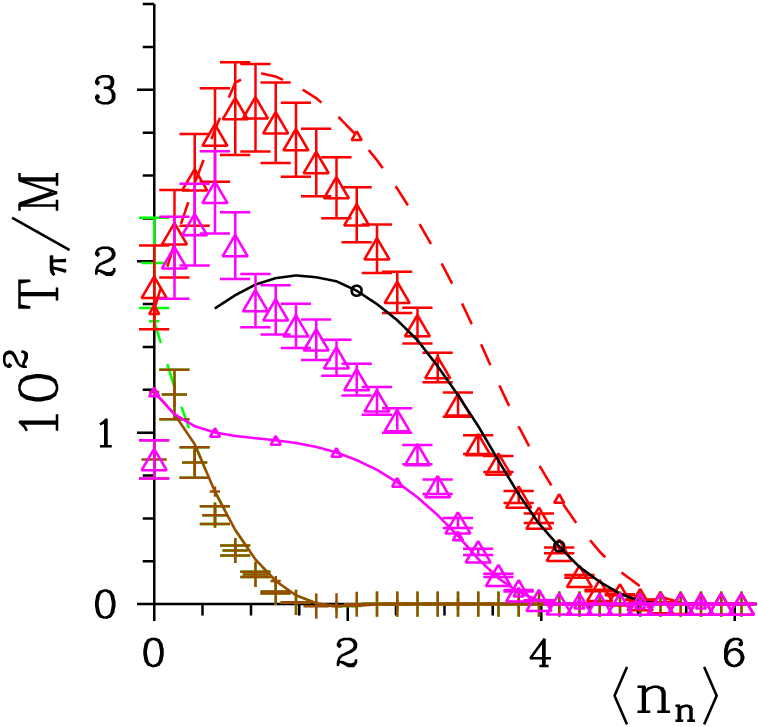}}
  \centerline{(e)\hspace{0.22\hsize} (f) \hspace{0.22\hsize}
    (g)\hspace{0.22\hsize} (h)}
 \caption{(a) [(e)] Seralian $ \Delta_2 $, (b) [(f)] 2-beam negativity per mode $ E_{N,2}/M $,     (c) [(g)] 3-beam negativity per mode $ E_{N,3}/M $, and
   (d) [(h)] cotangle per mode $ T_\pi/M $ for $ M=6.7 $ [$ M=40 $] modes as they
  depend on 1-beam mean noise photon number $ \langle n_{\rm n} \rangle $.
  Isolated symbols with error bars give experimental data, curves
  originate in the general Gaussian model.
  Black curves with $ \circ $ in (e---h) denote \emph{exact} values obtained
  from the intensity moments up to sixth order.
  Fourth-order moments provide the
  lower (green dashed curves and symbols +) and upper (red dashed curves and symbols $ \triangle $) bounds for quantum correlations. Direct estimation of the seralian $ \Delta_2 $ using fourth-order moments results in
  the lower (brown solid curves and symbols +) and upper (magenta solid curves and symbols $ \triangle $) bounds for quantum correlations. Lower bounds for $ E_{N,2}/M $,
  $ E_{N,3}/M $, and $ T_\pi/M $ are close or even coincide. Symbols and curves for $ \Delta_2 $ correspond to those for $ E_{N,2}/M $, $ E_{N,3}/M $, and $ T_\pi/M $.}
\label{fig10}
\end{figure*}

In the second, alternative, way the part of seralian $ \Delta_2 $
expressible directly in terms of intensity moments is computed and
the remaining part is estimated via its lower and upper bounds
(for details, see \cite{Sudak2025}). This more straightforward
approach limits more the allowed values of $ E_{N,2} $, $ E_{N,3}
$, and $ T_\pi $, as it is apparent from the curves in
Figs.~\ref{fig10}(b-d,f-h). Comparison of the curves in these
figures reveal that whereas the lower bounds from both approaches
are close or even identical, the upper bounds from the second
approach are considerably more restrictive.

Provided that the experimental fields are divided into $ M=6.7 $
independent modes in each beam, the second approach assigns
considerably lower values to the cotangle $ T_\pi $ than the first
approach and identifies the genuine 3-mode entanglement only in
the fields with $ \langle n_{\rm n} \rangle \le 1 $  [see
Fig.~\ref{fig10}(d)]. Compared to this, the use of intensity
moments up to sixth order reveals 3-beam genuine entanglement in
the fields with $ \langle n_{\rm n} \rangle \in \langle 1,2\rangle
$ [see Fig.~\ref{fig7}(c)]. However, the use of
sixth-order moments is rather demanding for experimental precision
and so it fails to identify the genuine 3-beam entanglement for $
\langle n_{\rm n} \rangle \le 1 $. This situation demonstrates the
importance of simpler approaches in the analysis of experimental
data and the need to perform the analysis in consecutive steps
starting with the simplest approach and gradually improving the
knowledge about the investigated states by applying more and more
elaborated approaches (that are more and more prone to the
experimental errors).

The comparison of results reached by different approaches depends
on the number $ M $ of independent modes in the beams of the
analyzed fields. We note that the beams' division into modes depends on how
we use the fields in measurement applications, information
processing, etc. Comparing the negativities $ E_{N,2} $ and $
E_{N,3} $ and cotangle $ T_\pi $ in Figs.~\ref{fig10}(b-d,f-h)
drawn for $ M=6.7 $ and $ M=40 $, we can see that the greater
numbers $ M $ of modes make the discrepancies of the results of
different approaches considerably smaller.
On the other side and considering experimental data, low
intensities in single-mode fields (greater number $ M $ of modes)
mean greater relative errors in determining their intensity
moments. These errors prevented us from determining the
experimental values of seralian $ \Delta_2 $ for $ M=40 $ in
Fig.~\ref{fig10}(e) and subsequently the corresponding values of
negativities $ E_{N,2} $ and $ E_{N,3} $ and cotangle $ T_\pi $ in
Figs.~\ref{fig10}(f-h) applying sixth-order intensity moments.

The comparison of curves in Figs.~\ref{fig10}(b-d,f-h)
demonstrates an important property of quantum correlations --
their dependence on the number $ M $ of independent modes that
constitute the fields. Whereas the genuine 3-beam entanglement is
identified in the fields with $ \langle n_{\rm n} \rangle \le 2 $
for $ M=6.7$, it occurs for a much broader group of fields with $
\langle n_{\rm n} \rangle \le 5 $ for $ M=40 $. This behavior
reflects specific influence of the noise when concealing the
entanglement of different forms: Dividing an overall field into a
greater number of independent modes effectively (partially)
prevents entanglement from the detrimental effects of the noise
\cite{Arkhipov2015}.

\section{Experimental errors and the role of the number of measurements}

\begin{figure*}  
  \centerline{\includegraphics[width=0.23\hsize]{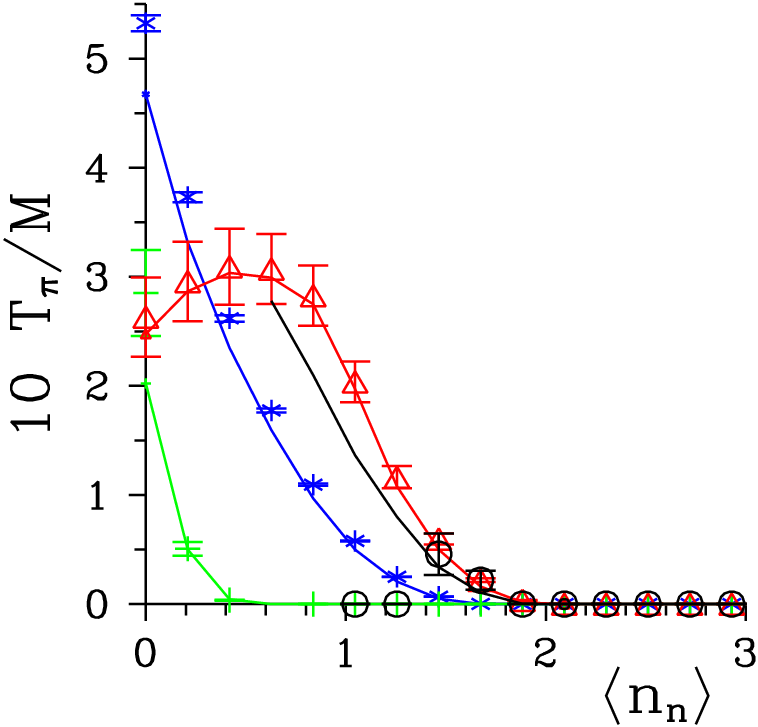}
    \hspace{2mm}
     \includegraphics[width=0.23\hsize]{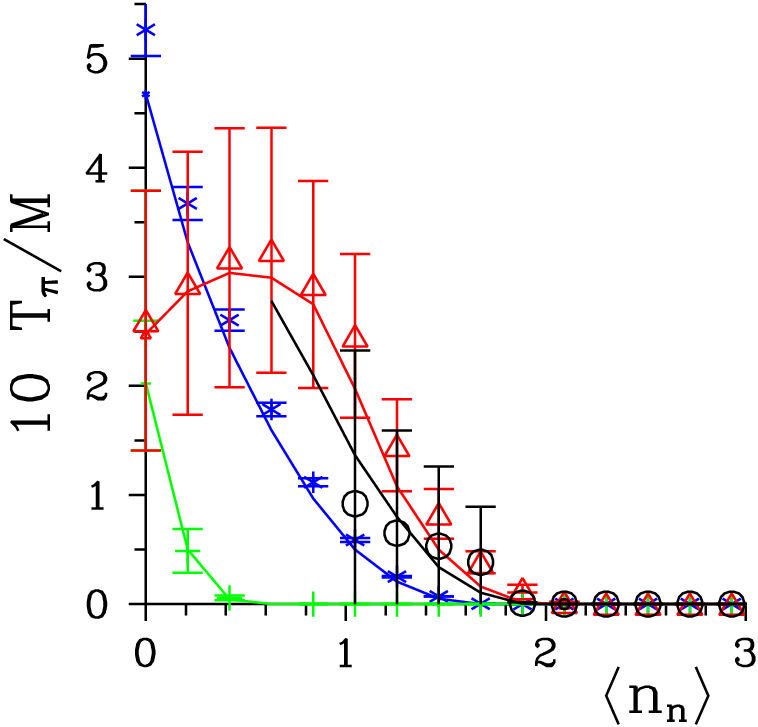}
    \hspace{2mm}
  \includegraphics[width=0.23\hsize]{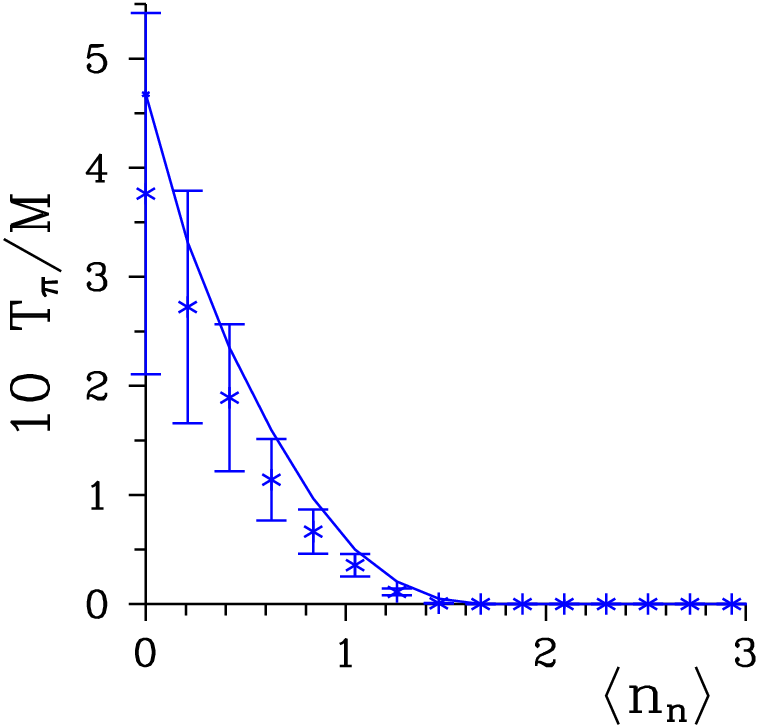} }
  \centerline{(a)\hspace{0.22\hsize} (b) \hspace{0.22\hsize}
   (c)}

 \caption{Cotangle per mode $ T_\pi/M $ for $ M=6.7 $ modes as they
  depend on 1-beam mean noise photon number $ \langle n_{\rm n} \rangle $
  determined from an ensemble containing (a) $ 1/6 \times 10^8 $, (b) $ 1/6 \times 10^7 $, and (c) $ 1/6 \times 10^5 $ measurement realizations. Symbols and curves are described in the caption to Fig.~\ref{fig5}.}
\label{fig11}
\end{figure*}

The quantities characterizing the experimental STBGSs were determined in three different ways that use in turn up
to sixth-order, fourth-order, and second-order intensity moments.
The use of intensity moments of different orders puts different requirements on the statistical precision of their determination derived from the number of measurement realizations. To estimate the needed
number of measurement repetitions, we draw in Fig.~\ref{fig11}
the cotangle per mode $ T_{\pi}/M $ determined for $ 1/6 \times
10^8 $, $ 1/6 \times 10^7/6 $, and $ 1/6 \times 10^5 $ measurement
realizations. Whereas the $ 6.95/6 \times 10^8 $ measurement
realizations used when drawing the cotangle per mode $ T_{\pi}/M $
in Fig.~\ref{fig7}(c) allows to reliably obtain the
sixth-order intensity moments and thus apply all three approaches
in analyzing the data, the reduced numbers of measurement
repetitions applied in Figs.~\ref{fig11}(a,b) suffice to reveal safely the intensity moments only up to the fourth order. Finally, the
$ 1/6 \times 10^5 $ measurement repetitions used when drawing the
graph in Fig.~\ref{fig11}(c) provide only second-order intensity
moments with sufficiently low experimental error: Nevertheless,
additional constraints on the form of the analyzed states
compensate for such low number of measurement repetitions and they allow to obtain the cotangle per mode $ T_{\pi}/M $ plotted in Fig.~\ref{fig11}(c).


\begin{thebibliography}{42}%
\makeatletter
\providecommand \@ifxundefined [1]{%
 \@ifx{#1\undefined}
}%
\providecommand \@ifnum [1]{%
 \ifnum #1\expandafter \@firstoftwo
 \else \expandafter \@secondoftwo
 \fi
}%
\providecommand \@ifx [1]{%
 \ifx #1\expandafter \@firstoftwo
 \else \expandafter \@secondoftwo
 \fi
}%
\providecommand \natexlab [1]{#1}%
\providecommand \enquote  [1]{``#1''}%
\providecommand \bibnamefont  [1]{#1}%
\providecommand \bibfnamefont [1]{#1}%
\providecommand \citenamefont [1]{#1}%
\providecommand \href@noop [0]{\@secondoftwo}%
\providecommand \href [0]{\begingroup \@sanitize@url \@href}%
\providecommand \@href[1]{\@@startlink{#1}\@@href}%
\providecommand \@@href[1]{\endgroup#1\@@endlink}%
\providecommand \@sanitize@url [0]{\catcode `\\12\catcode
`\$12\catcode
  `\&12\catcode `\#12\catcode `\^12\catcode `\_12\catcode `\%12\relax}%
\providecommand \@@startlink[1]{}%
\providecommand \@@endlink[0]{}%
\providecommand \url  [0]{\begingroup\@sanitize@url \@url }%
\providecommand \@url [1]{\endgroup\@href {#1}{\urlprefix }}%
\providecommand \urlprefix  [0]{URL }%
\providecommand \Eprint [0]{\href }%
\providecommand \doibase [0]{https://doi.org/}%
\providecommand \selectlanguage [0]{\@gobble}%
\providecommand \bibinfo  [0]{\@secondoftwo}%
\providecommand \bibfield  [0]{\@secondoftwo}%
\providecommand \translation [1]{[#1]}%
\providecommand \BibitemOpen [0]{}%
\providecommand \bibitemStop [0]{}%
\providecommand \bibitemNoStop [0]{.\EOS\space}%
\providecommand \EOS [0]{\spacefactor3000\relax}%
\providecommand \BibitemShut  [1]{\csname bibitem#1\endcsname}%
\let\auto@bib@innerbib\@empty
\bibitem [{\citenamefont {Stoler}(1970)}]{Stoler1970}%
  \BibitemOpen
  \bibfield  {author} {\bibinfo {author} {\bibfnamefont {D.}~\bibnamefont
  {Stoler}},\ }\bibfield  {title} {\bibinfo {title} {Equivalence classes of
  minimum uncertainty packets},\ }\href@noop {} {\bibfield  {journal} {\bibinfo
   {journal} {Phys. Rev. D}\ }\textbf {\bibinfo {volume} {1}},\ \bibinfo
  {pages} {3217} (\bibinfo {year} {1970})}\BibitemShut {NoStop}%
\bibitem [{\citenamefont {Slusher}\ \emph {et~al.}(1985)\citenamefont
  {Slusher}, \citenamefont {Hollberg}, \citenamefont {Yurke}, \citenamefont
  {Mertz},\ and\ \citenamefont {Valleys}}]{Slusher1985}%
  \BibitemOpen
  \bibfield  {author} {\bibinfo {author} {\bibfnamefont {R.~E.}\ \bibnamefont
  {Slusher}}, \bibinfo {author} {\bibfnamefont {L.~W.}\ \bibnamefont
  {Hollberg}}, \bibinfo {author} {\bibfnamefont {B.}~\bibnamefont {Yurke}},
  \bibinfo {author} {\bibfnamefont {J.~C.}\ \bibnamefont {Mertz}},\ and\
  \bibinfo {author} {\bibfnamefont {J.~F.}\ \bibnamefont {Valleys}},\
  }\bibfield  {title} {\bibinfo {title} {Observation of squeezed states
  generated by four-wave mixing in an optical cavity},\ }\href@noop {}
  {\bibfield  {journal} {\bibinfo  {journal} {Phys. Rev. Lett.}\ }\textbf
  {\bibinfo {volume} {55}},\ \bibinfo {pages} {2409} (\bibinfo {year}
  {1985})}\BibitemShut {NoStop}%
\bibitem [{\citenamefont {Dodonov}(2002)}]{Dodonov2002}%
  \BibitemOpen
  \bibfield  {author} {\bibinfo {author} {\bibfnamefont {V.~V.}\ \bibnamefont
  {Dodonov}},\ }\bibfield  {title} {\bibinfo {title} {Nonclassical states in
  quantum optics: A squeezed review of the first 75 years},\ }\href@noop {}
  {\bibfield  {journal} {\bibinfo  {journal} {J. Opt. B: Quantum Semiclass.
  Opt.}\ }\textbf {\bibinfo {volume} {4}},\ \bibinfo {pages} {R1} (\bibinfo
  {year} {2002})}\BibitemShut {NoStop}%
\bibitem [{\citenamefont {Laudenbach}\ \emph {et~al.}(1918)\citenamefont
  {Laudenbach}, \citenamefont {Pacher}, \citenamefont {Fung}, \citenamefont
  {Poppe}, \citenamefont {Peev}, \citenamefont {Schrenk}, \citenamefont
  {Hentschel},\ and\ \citenamefont {{H\" ubel}}}]{Laudenbach2018}%
  \BibitemOpen
  \bibfield  {author} {\bibinfo {author} {\bibfnamefont {F.}~\bibnamefont
  {Laudenbach}}, \bibinfo {author} {\bibfnamefont {C.}~\bibnamefont {Pacher}},
  \bibinfo {author} {\bibfnamefont {C.-H.~F.}\ \bibnamefont {Fung}}, \bibinfo
  {author} {\bibfnamefont {A.}~\bibnamefont {Poppe}}, \bibinfo {author}
  {\bibfnamefont {M.}~\bibnamefont {Peev}}, \bibinfo {author} {\bibfnamefont
  {B.}~\bibnamefont {Schrenk}}, \bibinfo {author} {\bibfnamefont
  {M.}~\bibnamefont {Hentschel}},\ and\ \bibinfo {author} {\bibfnamefont
  {P.~W.~H.}\ \bibnamefont {{H\" ubel}}},\ }\bibfield  {title} {\bibinfo
  {title} {Continuous-variable quantum key distribution with {Gaussian}
  modulation—the theory of practical implementations},\ }\href@noop {}
  {\bibfield  {journal} {\bibinfo  {journal} {Adv. Quantum Technol.}\ }\textbf
  {\bibinfo {volume} {1}},\ \bibinfo {pages} {1800011} (\bibinfo {year}
  {1918})}\BibitemShut {NoStop}%
\bibitem [{\citenamefont {Franzen}\ \emph {et~al.}(2006)\citenamefont
  {Franzen}, \citenamefont {Hage}, \citenamefont {DiGuglielmo}, \citenamefont
  {{Fiur\' a\v{s}ek}},\ and\ \citenamefont {Schnabel}}]{Franzen2006}%
  \BibitemOpen
  \bibfield  {author} {\bibinfo {author} {\bibfnamefont {A.}~\bibnamefont
  {Franzen}}, \bibinfo {author} {\bibfnamefont {B.}~\bibnamefont {Hage}},
  \bibinfo {author} {\bibfnamefont {J.}~\bibnamefont {DiGuglielmo}}, \bibinfo
  {author} {\bibfnamefont {J.}~\bibnamefont {{Fiur\' a\v{s}ek}}},\ and\
  \bibinfo {author} {\bibfnamefont {R.}~\bibnamefont {Schnabel}},\ }\bibfield
  {title} {\bibinfo {title} {Experimental demonstration of continuous variable
  purification of squeezed states},\ }\href@noop {} {\bibfield  {journal}
  {\bibinfo  {journal} {Phys. Rev. Lett.}\ }\textbf {\bibinfo {volume} {97}},\
  \bibinfo {pages} {150505} (\bibinfo {year} {2006})}\BibitemShut {NoStop}%
\bibitem [{\citenamefont {Grebien}\ \emph {et~al.}(1922)\citenamefont
  {Grebien}, \citenamefont {{G\" ottsch}}, \citenamefont {Hage}, \citenamefont
  {{Fiur\' a\v{s}ek}},\ and\ \citenamefont {Schnabel}}]{Grebien2022}%
  \BibitemOpen
  \bibfield  {author} {\bibinfo {author} {\bibfnamefont {S.}~\bibnamefont
  {Grebien}}, \bibinfo {author} {\bibfnamefont {J.}~\bibnamefont {{G\"
  ottsch}}}, \bibinfo {author} {\bibfnamefont {B.}~\bibnamefont {Hage}},
  \bibinfo {author} {\bibfnamefont {J.}~\bibnamefont {{Fiur\' a\v{s}ek}}},\
  and\ \bibinfo {author} {\bibfnamefont {R.}~\bibnamefont {Schnabel}},\
  }\bibfield  {title} {\bibinfo {title} {Multistep two-copy distillation of
  squeezed states via two-photon subtraction},\ }\href@noop {} {\bibfield
  {journal} {\bibinfo  {journal} {Phys. Rev. Lett.}\ }\textbf {\bibinfo
  {volume} {129}},\ \bibinfo {pages} {273604} (\bibinfo {year}
  {1922})}\BibitemShut {NoStop}%
\bibitem [{\citenamefont {Fukui}\ \emph {et~al.}(2018)\citenamefont {Fukui},
  \citenamefont {Tomita}, \citenamefont {Okamoto},\ and\ \citenamefont
  {Fujii}}]{FukuiPhysRevX8_2018}%
  \BibitemOpen
  \bibfield  {author} {\bibinfo {author} {\bibfnamefont {K.}~\bibnamefont
  {Fukui}}, \bibinfo {author} {\bibfnamefont {A.}~\bibnamefont {Tomita}},
  \bibinfo {author} {\bibfnamefont {A.}~\bibnamefont {Okamoto}},\ and\ \bibinfo
  {author} {\bibfnamefont {K.}~\bibnamefont {Fujii}},\ }\bibfield  {title}
  {\bibinfo {title} {High-threshold fault-tolerant quantum computation with
  analog quantum error correction},\ }\href
  {https://doi.org/10.1103/PhysRevX.8.021054} {\bibfield  {journal} {\bibinfo
  {journal} {Phys. Rev. X}\ }\textbf {\bibinfo {volume} {8}},\ \bibinfo {pages}
  {021054} (\bibinfo {year} {2018})}\BibitemShut {NoStop}%
\bibitem [{\citenamefont {Menicucci}(2014)}]{MenicucciPRL112_2014}%
  \BibitemOpen
  \bibfield  {author} {\bibinfo {author} {\bibfnamefont {N.~C.}\ \bibnamefont
  {Menicucci}},\ }\bibfield  {title} {\bibinfo {title} {Fault-tolerant
  measurement-based quantum computing with continuous-variable cluster
  states},\ }\href {https://doi.org/10.1103/PhysRevLett.112.120504} {\bibfield
  {journal} {\bibinfo  {journal} {Phys. Rev. Lett.}\ }\textbf {\bibinfo
  {volume} {112}},\ \bibinfo {pages} {120504} (\bibinfo {year}
  {2014})}\BibitemShut {NoStop}%
\bibitem [{\citenamefont {Schnabel}(2017)}]{Schnabel2017}%
  \BibitemOpen
  \bibfield  {author} {\bibinfo {author} {\bibfnamefont {R.}~\bibnamefont
  {Schnabel}},\ }\bibfield  {title} {\bibinfo {title} {Squeezed states of light
  and their applications in laser interferometers},\ }\href@noop {} {\bibfield
  {journal} {\bibinfo  {journal} {Phys. Rep.}\ }\textbf {\bibinfo {volume}
  {684}},\ \bibinfo {pages} {1} (\bibinfo {year} {2017})}\BibitemShut {NoStop}%
\bibitem [{\citenamefont {Lvovsky}\ and\ \citenamefont
  {Raymer}(2009)}]{Lvovsky2009}%
  \BibitemOpen
  \bibfield  {author} {\bibinfo {author} {\bibfnamefont {A.~I.}\ \bibnamefont
  {Lvovsky}}\ and\ \bibinfo {author} {\bibfnamefont {M.~G.}\ \bibnamefont
  {Raymer}},\ }\bibfield  {title} {\bibinfo {title} {Continuous-variable
  optical quantum-state tomography},\ }\href@noop {} {\bibfield  {journal}
  {\bibinfo  {journal} {Rev. Mod. Phys.}\ }\textbf {\bibinfo {volume} {81}},\
  \bibinfo {pages} {299} (\bibinfo {year} {2009})}\BibitemShut {NoStop}%
\bibitem [{\citenamefont {Mandel}\ and\ \citenamefont
  {Wolf}(1995)}]{Mandel1995}%
  \BibitemOpen
  \bibfield  {author} {\bibinfo {author} {\bibfnamefont {L.}~\bibnamefont
  {Mandel}}\ and\ \bibinfo {author} {\bibfnamefont {E.}~\bibnamefont {Wolf}},\
  }\href@noop {} {\emph {\bibinfo {title} {Optical Coherence and Quantum
  Optics}}}\ (\bibinfo  {publisher} {Cambridge Univ. Press, Cambridge},\
  \bibinfo {year} {1995})\BibitemShut {NoStop}%
\bibitem [{\citenamefont {Sudak}\ \emph {et~al.}(2025)\citenamefont {Sudak},
  \citenamefont {{Pe\v{r}ina~Jr.}}, \citenamefont {Barasi\'nski},\ and\
  \citenamefont {\v{C}ernoch}}]{Sudak2025}%
  \BibitemOpen
  \bibfield  {author} {\bibinfo {author} {\bibfnamefont {N.}~\bibnamefont
  {Sudak}}, \bibinfo {author} {\bibfnamefont {J.}~\bibnamefont
  {{Pe\v{r}ina~Jr.}}}, \bibinfo {author} {\bibfnamefont {A.}~\bibnamefont
  {Barasi\'nski}},\ and\ \bibinfo {author} {\bibfnamefont {A.}~\bibnamefont
  {\v{C}ernoch}},\ }\bibfield  {title} {\bibinfo {title} {Efficient
  characterization of {N}-beam {Gaussian} fields through photon-number
  measurements: Quantum universal invariants},\ }\href@noop {} {\bibfield
  {journal} {\bibinfo  {journal} {Phys. Rev. Res.}\ }\textbf {\bibinfo {volume}
  {7}},\ \bibinfo {pages} {023278} (\bibinfo {year} {2025})}\BibitemShut
  {NoStop}%
\bibitem [{\citenamefont {Adesso}\ \emph
  {et~al.}(2004{\natexlab{a}})\citenamefont {Adesso}, \citenamefont
  {Serafini},\ and\ \citenamefont {Illuminati}}]{Adesso2004}%
  \BibitemOpen
  \bibfield  {author} {\bibinfo {author} {\bibfnamefont {G.}~\bibnamefont
  {Adesso}}, \bibinfo {author} {\bibfnamefont {A.}~\bibnamefont {Serafini}},\
  and\ \bibinfo {author} {\bibfnamefont {F.}~\bibnamefont {Illuminati}},\
  }\bibfield  {title} {\bibinfo {title} {Quantification and scaling of
  multipartite entanglement in continuous variable systems},\ }\href
  {https://doi.org/10.1103/PhysRevLett.93.220504} {\bibfield  {journal}
  {\bibinfo  {journal} {Phys. Rev. Lett.}\ }\textbf {\bibinfo {volume} {93}},\
  \bibinfo {pages} {220504} (\bibinfo {year} {2004}{\natexlab{a}})}\BibitemShut
  {NoStop}%
\bibitem [{\citenamefont {Adesso}\ \emph {et~al.}(2006)\citenamefont {Adesso},
  \citenamefont {Serafini},\ and\ \citenamefont
  {Illuminati}}]{Adessopra73_2006}%
  \BibitemOpen
  \bibfield  {author} {\bibinfo {author} {\bibfnamefont {G.}~\bibnamefont
  {Adesso}}, \bibinfo {author} {\bibfnamefont {A.}~\bibnamefont {Serafini}},\
  and\ \bibinfo {author} {\bibfnamefont {F.}~\bibnamefont {Illuminati}},\
  }\bibfield  {title} {\bibinfo {title} {Multipartite entanglement in
  three-mode {Gaussian} states of continuous-variable systems: Quantification,
  sharing structure, and decoherence},\ }\href@noop {} {\bibfield  {journal}
  {\bibinfo  {journal} {Phys. Rev. A}\ }\textbf {\bibinfo {volume} {73}},\
  \bibinfo {pages} {032345} (\bibinfo {year} {2006})}\BibitemShut {NoStop}%
\bibitem [{\citenamefont {Pirandola}\ \emph {et~al.}(2003)\citenamefont
  {Pirandola}, \citenamefont {Mancini}, \citenamefont {Vitali},\ and\
  \citenamefont {Tombesi}}]{PirandolaPRA68_2003}%
  \BibitemOpen
  \bibfield  {author} {\bibinfo {author} {\bibfnamefont {S.}~\bibnamefont
  {Pirandola}}, \bibinfo {author} {\bibfnamefont {S.}~\bibnamefont {Mancini}},
  \bibinfo {author} {\bibfnamefont {D.}~\bibnamefont {Vitali}},\ and\ \bibinfo
  {author} {\bibfnamefont {P.}~\bibnamefont {Tombesi}},\ }\bibfield  {title}
  {\bibinfo {title} {Continuous-variable entanglement and quantum-state
  teleportation between optical and macroscopic vibrational modes through
  radiation pressure},\ }\href {https://doi.org/10.1103/PhysRevA.68.062317}
  {\bibfield  {journal} {\bibinfo  {journal} {Phys. Rev. A}\ }\textbf {\bibinfo
  {volume} {68}},\ \bibinfo {pages} {062317} (\bibinfo {year}
  {2003})}\BibitemShut {NoStop}%
\bibitem [{\citenamefont {Yonezawa}\ \emph {et~al.}(2004)\citenamefont
  {Yonezawa}, \citenamefont {Aoki},\ and\ \citenamefont
  {Furusawa}}]{Yonezawanature431_2004}%
  \BibitemOpen
  \bibfield  {author} {\bibinfo {author} {\bibfnamefont {H.}~\bibnamefont
  {Yonezawa}}, \bibinfo {author} {\bibfnamefont {T.}~\bibnamefont {Aoki}},\
  and\ \bibinfo {author} {\bibfnamefont {A.}~\bibnamefont {Furusawa}},\
  }\bibfield  {title} {\bibinfo {title} {Demonstration of a quantum
  teleportation network for continuous variables},\ }\href@noop {} {\bibfield
  {journal} {\bibinfo  {journal} {Nature}\ }\textbf {\bibinfo {volume} {431}},\
  \bibinfo {pages} {430} (\bibinfo {year} {2004})}\BibitemShut {NoStop}%
\bibitem [{\citenamefont {Ferraro}\ and\ \citenamefont
  {Paris}(2005)}]{FerraroPRA72_2005}%
  \BibitemOpen
  \bibfield  {author} {\bibinfo {author} {\bibfnamefont {A.}~\bibnamefont
  {Ferraro}}\ and\ \bibinfo {author} {\bibfnamefont {M.~G.~A.}\ \bibnamefont
  {Paris}},\ }\bibfield  {title} {\bibinfo {title} {Multimode entanglement and
  telecloning in a noisy environment},\ }\href
  {https://doi.org/10.1103/PhysRevA.72.032312} {\bibfield  {journal} {\bibinfo
  {journal} {Phys. Rev. A}\ }\textbf {\bibinfo {volume} {72}},\ \bibinfo
  {pages} {032312} (\bibinfo {year} {2005})}\BibitemShut {NoStop}%
\bibitem [{\citenamefont {Barasi\ifmmode~\acute{n}\else \'{n}\fi{}ski}\ \emph
  {et~al.}(2023)\citenamefont {Barasi\ifmmode~\acute{n}\else \'{n}\fi{}ski},
  \citenamefont {Pe\ifmmode~\check{r}\else \v{r}\fi{}ina},\ and\ \citenamefont
  {\ifmmode~\check{C}\else \v{C}\fi{}ernoch}}]{BarasinskiPRL2023}%
  \BibitemOpen
  \bibfield  {author} {\bibinfo {author} {\bibfnamefont {A.}~\bibnamefont
  {Barasi\ifmmode~\acute{n}\else \'{n}\fi{}ski}}, \bibinfo {author}
  {\bibfnamefont {J.}~\bibnamefont {Pe\ifmmode~\check{r}\else \v{r}\fi{}ina}},\
  and\ \bibinfo {author} {\bibfnamefont {A.}~\bibnamefont
  {\ifmmode~\check{C}\else \v{C}\fi{}ernoch}},\ }\bibfield  {title} {\bibinfo
  {title} {Quantification of quantum correlations in two-beam {Gaussian} states
  using photon-number measurements},\ }\href
  {https://doi.org/10.1103/PhysRevLett.130.043603} {\bibfield  {journal}
  {\bibinfo  {journal} {Phys. Rev. Lett.}\ }\textbf {\bibinfo {volume} {130}},\
  \bibinfo {pages} {043603} (\bibinfo {year} {2023})}\BibitemShut {NoStop}%
\bibitem [{\citenamefont {Adesso}\ and\ \citenamefont
  {Illuminati}(2006)}]{Adesso_NJP2006}%
  \BibitemOpen
  \bibfield  {author} {\bibinfo {author} {\bibfnamefont {G.}~\bibnamefont
  {Adesso}}\ and\ \bibinfo {author} {\bibfnamefont {F.}~\bibnamefont
  {Illuminati}},\ }\bibfield  {title} {\bibinfo {title} {Continuous variable
  tangle, monogamy inequality, and entanglement sharing in {Gaussian} states of
  continuous variable systems},\ }\href
  {https://doi.org/10.1088/1367-2630/8/1/015} {\bibfield  {journal} {\bibinfo
  {journal} {New Journal of Physics}\ }\textbf {\bibinfo {volume} {8}},\
  \bibinfo {pages} {15} (\bibinfo {year} {2006})}\BibitemShut {NoStop}%
\bibitem [{\citenamefont {{Pe\v{r}ina~Jr.}}\ \emph {et~al.}(2021)\citenamefont
  {{Pe\v{r}ina~Jr.}}, \citenamefont {\v{C}ernoch},\ and\ \citenamefont
  {Soubusta}}]{PerinaJr.2021}%
  \BibitemOpen
  \bibfield  {author} {\bibinfo {author} {\bibfnamefont {J.}~\bibnamefont
  {{Pe\v{r}ina~Jr.}}}, \bibinfo {author} {\bibfnamefont {A.}~\bibnamefont
  {\v{C}ernoch}},\ and\ \bibinfo {author} {\bibfnamefont {J.}~\bibnamefont
  {Soubusta}},\ }\bibfield  {title} {\bibinfo {title} {Compound twin beams
  without the need of genuine photon-number-resolving detection},\ }\href@noop
  {} {\bibfield  {journal} {\bibinfo  {journal} {Phys. Rev. Applied}\ }\textbf
  {\bibinfo {volume} {16}},\ \bibinfo {pages} {024061} (\bibinfo {year}
  {2021})}\BibitemShut {NoStop}%
\bibitem [{\citenamefont {Giedke}\ \emph {et~al.}(2001)\citenamefont {Giedke},
  \citenamefont {Kraus}, \citenamefont {Lewenstein},\ and\ \citenamefont
  {Cirac}}]{GiedkePPA64_2001}%
  \BibitemOpen
  \bibfield  {author} {\bibinfo {author} {\bibfnamefont {G.}~\bibnamefont
  {Giedke}}, \bibinfo {author} {\bibfnamefont {B.}~\bibnamefont {Kraus}},
  \bibinfo {author} {\bibfnamefont {M.}~\bibnamefont {Lewenstein}},\ and\
  \bibinfo {author} {\bibfnamefont {J.~I.}\ \bibnamefont {Cirac}},\ }\bibfield
  {title} {\bibinfo {title} {Separability properties of three-mode {Gaussian}
  states},\ }\href {https://doi.org/10.1103/PhysRevA.64.052303} {\bibfield
  {journal} {\bibinfo  {journal} {Phys. Rev. A}\ }\textbf {\bibinfo {volume}
  {64}},\ \bibinfo {pages} {052303} (\bibinfo {year} {2001})}\BibitemShut
  {NoStop}%
\bibitem [{com({\natexlab{a}})}]{comment}%
  \BibitemOpen
  \href@noop {} {} ({\natexlab{a}}),\ \bibinfo {note} {in the literature, such
  tripartite CV Gaussian states are commonly referred to as three-mode Gaussian
  states. However, to account for the multimode structure of the experimental
  fields, we adopt the term three-beam symmetric Gaussian states to prevent any
  potential misunderstanding.}\BibitemShut {Stop}%
\bibitem [{\citenamefont {Robertson}(1929)}]{RobertsonPR34_1929}%
  \BibitemOpen
  \bibfield  {author} {\bibinfo {author} {\bibfnamefont {H.~P.}\ \bibnamefont
  {Robertson}},\ }\bibfield  {title} {\bibinfo {title} {The uncertainty
  principle},\ }\href {https://doi.org/10.1103/PhysRev.34.163} {\bibfield
  {journal} {\bibinfo  {journal} {Phys. Rev.}\ }\textbf {\bibinfo {volume}
  {34}},\ \bibinfo {pages} {163} (\bibinfo {year} {1929})}\BibitemShut
  {NoStop}%
\bibitem [{\citenamefont {Schrödinger}(1930)}]{Schrodinger1930}%
  \BibitemOpen
  \bibfield  {author} {\bibinfo {author} {\bibfnamefont {E.}~\bibnamefont
  {Schrödinger}},\ }\bibfield  {title} {\bibinfo {title} {Zum
  {Heisenbergschen} {Unsch\"{a}rfeprinzip}},\ }\href@noop {} {\bibfield
  {journal} {\bibinfo  {journal} {Ber. Kgl. Akad. Wiss. Berlin}\ }\textbf
  {\bibinfo {volume} {24}},\ \bibinfo {pages} {296} (\bibinfo {year}
  {1930})}\BibitemShut {NoStop}%
\bibitem [{\citenamefont {Serafini}(2006)}]{SerafiniPRL96_2006}%
  \BibitemOpen
  \bibfield  {author} {\bibinfo {author} {\bibfnamefont {A.}~\bibnamefont
  {Serafini}},\ }\bibfield  {title} {\bibinfo {title} {Multimode uncertainty
  relations and separability of continuous variable states},\ }\href
  {https://doi.org/10.1103/PhysRevLett.96.110402} {\bibfield  {journal}
  {\bibinfo  {journal} {Phys. Rev. Lett.}\ }\textbf {\bibinfo {volume} {96}},\
  \bibinfo {pages} {110402} (\bibinfo {year} {2006})}\BibitemShut {NoStop}%
\bibitem [{\citenamefont {Simon}(2000)}]{Simon2000}%
  \BibitemOpen
  \bibfield  {author} {\bibinfo {author} {\bibfnamefont {R.}~\bibnamefont
  {Simon}},\ }\bibfield  {title} {\bibinfo {title} {{Peres-Horodecki}
  separability criterion for continuous variable systems},\ }\href@noop {}
  {\bibfield  {journal} {\bibinfo  {journal} {Phys. Rev. Lett.}\ }\textbf
  {\bibinfo {volume} {84}},\ \bibinfo {pages} {2726} (\bibinfo {year}
  {2000})}\BibitemShut {NoStop}%
\bibitem [{\citenamefont {Duan}\ \emph {et~al.}(2000)\citenamefont {Duan},
  \citenamefont {Giedke}, \citenamefont {Cirac},\ and\ \citenamefont
  {Zoller}}]{DuanPRL84_2000}%
  \BibitemOpen
  \bibfield  {author} {\bibinfo {author} {\bibfnamefont {L.-M.}\ \bibnamefont
  {Duan}}, \bibinfo {author} {\bibfnamefont {G.}~\bibnamefont {Giedke}},
  \bibinfo {author} {\bibfnamefont {J.~I.}\ \bibnamefont {Cirac}},\ and\
  \bibinfo {author} {\bibfnamefont {P.}~\bibnamefont {Zoller}},\ }\bibfield
  {title} {\bibinfo {title} {Inseparability criterion for continuous variable
  systems},\ }\href {https://doi.org/10.1103/PhysRevLett.84.2722} {\bibfield
  {journal} {\bibinfo  {journal} {Phys. Rev. Lett.}\ }\textbf {\bibinfo
  {volume} {84}},\ \bibinfo {pages} {2722} (\bibinfo {year}
  {2000})}\BibitemShut {NoStop}%
\bibitem [{\citenamefont {Serafini}\ \emph {et~al.}(2005)\citenamefont
  {Serafini}, \citenamefont {Adesso},\ and\ \citenamefont
  {Illuminati}}]{SerafiniPRA71_2005}%
  \BibitemOpen
  \bibfield  {author} {\bibinfo {author} {\bibfnamefont {A.}~\bibnamefont
  {Serafini}}, \bibinfo {author} {\bibfnamefont {G.}~\bibnamefont {Adesso}},\
  and\ \bibinfo {author} {\bibfnamefont {F.}~\bibnamefont {Illuminati}},\
  }\bibfield  {title} {\bibinfo {title} {Unitarily localizable entanglement of
  {Gaussian} states},\ }\href {https://doi.org/10.1103/PhysRevA.71.032349}
  {\bibfield  {journal} {\bibinfo  {journal} {Phys. Rev. A}\ }\textbf {\bibinfo
  {volume} {71}},\ \bibinfo {pages} {032349} (\bibinfo {year}
  {2005})}\BibitemShut {NoStop}%
\bibitem [{\citenamefont {Pe\v{r}ina}(1991)}]{Perina1991}%
  \BibitemOpen
  \bibfield  {author} {\bibinfo {author} {\bibfnamefont {J.}~\bibnamefont
  {Pe\v{r}ina}},\ }\href@noop {} {\emph {\bibinfo {title} {Quantum Statistics
  of Linear and Nonlinear Optical Phenomena}}}\ (\bibinfo  {publisher} {Kluwer,
  Dordrecht},\ \bibinfo {year} {1991})\BibitemShut {NoStop}%
\bibitem [{\citenamefont {Adesso}\ \emph
  {et~al.}(2004{\natexlab{b}})\citenamefont {Adesso}, \citenamefont
  {Serafini},\ and\ \citenamefont {Illuminati}}]{Adesso2004b}%
  \BibitemOpen
  \bibfield  {author} {\bibinfo {author} {\bibfnamefont {G.}~\bibnamefont
  {Adesso}}, \bibinfo {author} {\bibfnamefont {A.}~\bibnamefont {Serafini}},\
  and\ \bibinfo {author} {\bibfnamefont {F.}~\bibnamefont {Illuminati}},\
  }\bibfield  {title} {\bibinfo {title} {Determination of continuous variable
  entanglement by purity measurements},\ }\href@noop {} {\bibfield  {journal}
  {\bibinfo  {journal} {Phys. Rev. Lett.}\ }\textbf {\bibinfo {volume} {92}},\
  \bibinfo {pages} {087901} (\bibinfo {year} {2004}{\natexlab{b}})}\BibitemShut
  {NoStop}%
\bibitem [{\citenamefont {Adesso}\ \emph {et~al.}(2007)\citenamefont {Adesso},
  \citenamefont {Serafini},\ and\ \citenamefont {Illuminati}}]{AdessoNPJ2007}%
  \BibitemOpen
  \bibfield  {author} {\bibinfo {author} {\bibfnamefont {G.}~\bibnamefont
  {Adesso}}, \bibinfo {author} {\bibfnamefont {A.}~\bibnamefont {Serafini}},\
  and\ \bibinfo {author} {\bibfnamefont {F.}~\bibnamefont {Illuminati}},\
  }\bibfield  {title} {\bibinfo {title} {Optical state engineering, quantum
  communication, and robustness of entanglement promiscuity in three-mode
  {Gaussian} states},\ }\href {https://doi.org/10.1088/1367-2630/9/3/060}
  {\bibfield  {journal} {\bibinfo  {journal} {New J. Phys.}\ }\textbf {\bibinfo
  {volume} {9}},\ \bibinfo {pages} {60} (\bibinfo {year} {2007})}\BibitemShut
  {NoStop}%
\bibitem [{\citenamefont {Chen}(2005)}]{CHEN2005121}%
  \BibitemOpen
  \bibfield  {author} {\bibinfo {author} {\bibfnamefont {X.-Y.}\ \bibnamefont
  {Chen}},\ }\bibfield  {title} {\bibinfo {title} {Entanglement bounds of
  tripartite squeezed thermal states},\ }\href
  {https://doi.org/https://doi.org/10.1016/j.physleta.2004.12.025} {\bibfield
  {journal} {\bibinfo  {journal} {Phys. Lett. A}\ }\textbf {\bibinfo {volume}
  {335}},\ \bibinfo {pages} {121} (\bibinfo {year} {2005})}\BibitemShut
  {NoStop}%
\bibitem [{\citenamefont {Hill}\ and\ \citenamefont
  {Wootters}(1997)}]{Hill1997}%
  \BibitemOpen
  \bibfield  {author} {\bibinfo {author} {\bibfnamefont {S.~A.}\ \bibnamefont
  {Hill}}\ and\ \bibinfo {author} {\bibfnamefont {W.~K.}\ \bibnamefont
  {Wootters}},\ }\bibfield  {title} {\bibinfo {title} {Computable
  entanglement},\ }\href@noop {} {\bibfield  {journal} {\bibinfo  {journal}
  {Phys. Rev. Lett.}\ }\textbf {\bibinfo {volume} {78}},\ \bibinfo {pages}
  {5022} (\bibinfo {year} {1997})}\BibitemShut {NoStop}%
\bibitem [{\citenamefont {Horodecki}\ \emph {et~al.}(2009)\citenamefont
  {Horodecki}, \citenamefont {Horodecki}, \citenamefont {Horodecki},\ and\
  \citenamefont {Horodecki}}]{Horodecki2009}%
  \BibitemOpen
  \bibfield  {author} {\bibinfo {author} {\bibfnamefont {R.}~\bibnamefont
  {Horodecki}}, \bibinfo {author} {\bibfnamefont {P.}~\bibnamefont
  {Horodecki}}, \bibinfo {author} {\bibfnamefont {M.}~\bibnamefont
  {Horodecki}},\ and\ \bibinfo {author} {\bibfnamefont {K.}~\bibnamefont
  {Horodecki}},\ }\bibfield  {title} {\bibinfo {title} {Quantum entanglement},\
  }\href@noop {} {\bibfield  {journal} {\bibinfo  {journal} {Rev. Mod. Phys.}\
  }\textbf {\bibinfo {volume} {81}},\ \bibinfo {pages} {865} (\bibinfo {year}
  {2009})}\BibitemShut {NoStop}%
\bibitem [{\citenamefont {Adesso}\ and\ \citenamefont
  {Illuminati}(2007)}]{Adesso2007}%
  \BibitemOpen
  \bibfield  {author} {\bibinfo {author} {\bibfnamefont {G.}~\bibnamefont
  {Adesso}}\ and\ \bibinfo {author} {\bibfnamefont {F.}~\bibnamefont
  {Illuminati}},\ }\bibfield  {title} {\bibinfo {title} {Entanglement in
  continuous-variable systems: recent advances and current perspectives},\
  }\href@noop {} {\bibfield  {journal} {\bibinfo  {journal} {J. Phys. A}\
  }\textbf {\bibinfo {volume} {40}},\ \bibinfo {pages} {7821} (\bibinfo {year}
  {2007})}\BibitemShut {NoStop}%
\bibitem [{\citenamefont {Kogias}\ \emph {et~al.}(2015)\citenamefont {Kogias},
  \citenamefont {Lee}, \citenamefont {Ragy},\ and\ \citenamefont
  {Adesso}}]{Kogias2015}%
  \BibitemOpen
  \bibfield  {author} {\bibinfo {author} {\bibfnamefont {I.}~\bibnamefont
  {Kogias}}, \bibinfo {author} {\bibfnamefont {A.~R.}\ \bibnamefont {Lee}},
  \bibinfo {author} {\bibfnamefont {S.}~\bibnamefont {Ragy}},\ and\ \bibinfo
  {author} {\bibfnamefont {G.}~\bibnamefont {Adesso}},\ }\bibfield  {title}
  {\bibinfo {title} {Quantification of {Gaussian} quantum steering},\
  }\href@noop {} {\bibfield  {journal} {\bibinfo  {journal} {Phys. Rev. Lett.}\
  }\textbf {\bibinfo {volume} {114}},\ \bibinfo {pages} {060403} (\bibinfo
  {year} {2015})}\BibitemShut {NoStop}%
\bibitem [{com({\natexlab{b}})}]{comment2}%
  \BibitemOpen
  \href@noop {} {} ({\natexlab{b}}),\ \bibinfo {note} {for STBGSs the Schur
  complement is given by $M_\sigma = \begin{pmatrix} \boldsymbol{\alpha} &
  \boldsymbol{\gamma}\\ \boldsymbol{\gamma} & \boldsymbol{\alpha} \end{pmatrix}
  - \begin{pmatrix} \boldsymbol{\gamma}\\ \boldsymbol{\gamma} \end{pmatrix}
  \boldsymbol{\alpha}^{-1} \begin{pmatrix} \boldsymbol{\gamma}&
  \boldsymbol{\gamma}\end{pmatrix}$}\BibitemShut {NoStop}%
\bibitem [{SM()}]{SM}%
  \BibitemOpen
  \href@noop {} {}\bibinfo {note} {See Supplemental Material at for the
  structure of the analyzed experimental multi-mode fields, experimental setup,
  reconstruction of experimental photocount histograms, multi-mode Gaussian
  model, and relation between multi-mode and single-mode intensity moments of
  3-beam Gaussian fields}\BibitemShut {NoStop}%
\bibitem [{\citenamefont {Dempster}\ \emph {et~al.}(1977)\citenamefont
  {Dempster}, \citenamefont {Laird},\ and\ \citenamefont
  {Rubin}}]{Dempster1977}%
  \BibitemOpen
  \bibfield  {author} {\bibinfo {author} {\bibfnamefont {A.~P.}\ \bibnamefont
  {Dempster}}, \bibinfo {author} {\bibfnamefont {N.~M.}\ \bibnamefont
  {Laird}},\ and\ \bibinfo {author} {\bibfnamefont {D.~B.}\ \bibnamefont
  {Rubin}},\ }\bibfield  {title} {\bibinfo {title} {Maximum likelihood from
  incomplete data via the {EM} algorithm},\ }\href@noop {} {\bibfield
  {journal} {\bibinfo  {journal} {J. Royal Statist. Soc. B}\ }\textbf {\bibinfo
  {volume} {39}},\ \bibinfo {pages} {1} (\bibinfo {year} {1977})}\BibitemShut
  {NoStop}%
\bibitem [{\citenamefont {Vardi}\ and\ \citenamefont {Lee}(1993)}]{Vardi1993}%
  \BibitemOpen
  \bibfield  {author} {\bibinfo {author} {\bibfnamefont {Y.}~\bibnamefont
  {Vardi}}\ and\ \bibinfo {author} {\bibfnamefont {D.}~\bibnamefont {Lee}},\
  }\bibfield  {title} {\bibinfo {title} {From image deblurring to optimal
  investments: Maximum likelihood solutions for positive linear inverse
  problems},\ }\href@noop {} {\bibfield  {journal} {\bibinfo  {journal} {J.
  Royal Statist. Soc. B}\ }\textbf {\bibinfo {volume} {55}},\ \bibinfo {pages}
  {569} (\bibinfo {year} {1993})}\BibitemShut {NoStop}%
\bibitem [{\citenamefont {Arkhipov}\ \emph {et~al.}(2015)\citenamefont
  {Arkhipov}, \citenamefont {{Pe\v{r}ina~Jr.}}, \citenamefont {Pe\v{r}ina},\
  and\ \citenamefont {Miranowicz}}]{Arkhipov2015}%
  \BibitemOpen
  \bibfield  {author} {\bibinfo {author} {\bibfnamefont {I.~I.}\ \bibnamefont
  {Arkhipov}}, \bibinfo {author} {\bibfnamefont {J.}~\bibnamefont
  {{Pe\v{r}ina~Jr.}}}, \bibinfo {author} {\bibfnamefont {J.}~\bibnamefont
  {Pe\v{r}ina}},\ and\ \bibinfo {author} {\bibfnamefont {A.}~\bibnamefont
  {Miranowicz}},\ }\bibfield  {title} {\bibinfo {title} {Comparative study of
  nonclassicality, entanglement, and dimensionality of multimode noisy twin
  beams},\ }\href@noop {} {\bibfield  {journal} {\bibinfo  {journal} {Phys.
  Rev. A}\ }\textbf {\bibinfo {volume} {91}},\ \bibinfo {pages} {033837}
  (\bibinfo {year} {2015})}\BibitemShut {NoStop}%
\bibitem [{\citenamefont {{Mich\'{a}lek}}\ \emph {et~al.}(2020)\citenamefont
  {{Mich\'{a}lek}}, \citenamefont {{Pe\v{r}ina~Jr.}},\ and\ \citenamefont
  {Haderka}}]{Michalek2020}%
  \BibitemOpen
  \bibfield  {author} {\bibinfo {author} {\bibfnamefont {V.}~\bibnamefont
  {{Mich\'{a}lek}}}, \bibinfo {author} {\bibfnamefont {J.}~\bibnamefont
  {{Pe\v{r}ina~Jr.}}},\ and\ \bibinfo {author} {\bibfnamefont {O.}~\bibnamefont
  {Haderka}},\ }\bibfield  {title} {\bibinfo {title} {Experimental
  quantification of the entanglement of noisy twin beams},\ }\href@noop {}
  {\bibfield  {journal} {\bibinfo  {journal} {Phys. Rev. Applied}\ }\textbf
  {\bibinfo {volume} {14}},\ \bibinfo {pages} {024003} (\bibinfo {year}
  {2020})}\BibitemShut {NoStop}%
\end{thebibliography}

%

\end{document}